\documentclass{aa}
\usepackage{amsmath,graphicx}
\usepackage{epsfig}
\usepackage{txfonts}
\usepackage{natbib}
\usepackage{rotating}
\bibpunct{(}{)}{;}{a}{}{,} 
\newcommand{\BA}{\tens{A}}

\newcommand{\Bs}{\vec{s}}

\newcommand{\Bn}{\vec{n}}

\newcommand{\Bd}{\vec{d}}

\newcommand{\Bx}{\vec{x}}

\newcommand{\bfref}{}

\graphicspath{{Figs/Finalfigs/}} 

\def\draftfig{false} 

\DeclareGraphicsExtensions{.eps, .ps}

\title{Component separation methods for the {\sc Planck} mission}
\authorrunning{S.~M.~Leach et al}
 \author{
S.~M.~Leach\inst{1,2},
J.-F.~Cardoso\inst{3,4},
C.~Baccigalupi\inst{1,2},
R.~B.~Barreiro\inst{5},
M.~Betoule\inst{3},
J.~Bobin\inst{6},
A.~Bonaldi\inst{7,8},
J.~Delabrouille\inst{3},
G.~de Zotti\inst{7,1},
C.~Dickinson\inst{9},
H.~K.~Eriksen\inst{10,11},
J.~Gonz{\'a}lez-Nuevo\inst{1},
F.~K.~Hansen\inst{10,11},
D.~Herranz\inst{5},
M.~Le~Jeune\inst{3},
M.~L{\'o}pez-Caniego\inst{12},
E.~Mart{\'\i}nez-Gonz\'alez\inst{5},
M.~Massardi\inst{1},
J.-B.~Melin\inst{13},
M.-A.~Miville-Desch\^{e}nes\inst{14},
G.~Patanchon\inst{3},
S.~Prunet\inst{15},
S.~Ricciardi\inst{7,16},
E.~Salerno\inst{17},
J.~L.~Sanz\inst{5},
J.-L.~Starck\inst{6},
F.~Stivoli\inst{1,2},
V.~Stolyarov\inst{12},
R.~Stompor\inst{3}\and
P.~Vielva\inst{5} 
}
\institute{
 SISSA-ISAS, Astrophysics Sector, via Beirut 4, Trieste, 34014, Italy
 \and
 INFN, Sezione di Trieste, Via Valerio 2, I-34014 Trieste, Italy
 \and
 CNRS \& Universit\'e Paris 7, Laboratoire APC, 10 rue A. Domon et L. Duquet, 75205 Paris Cedex 13, France
 \and
 Laboratoire de Traitement et Communication de l'Information (CNRS and Telecom ParisTech),
 46, rue Barrault, 75634 Paris, France
 \and
 Instituto de F\'{i}sica de Cantabria (CSIC-UC), Avda. de los Castros s/n, 39005 Santander, Spain
 \and
 CEA - Saclay, SEDI/Service d'Astrophysique, 91191 Gif-Sur-Yvette , France
 \and
 INAF-Osservatorio Astronomico di Padova, vicolo dell'Osservatorio 5, I-35122 Padova, Italy
 \and
 Dipartimento di Astronomia, vicolo dell'Osservatorio 5, I-35122 Padova, Italy
 \and
 Infrared Processing and Analysis Center, California Institute of Technology, M/S 220-6, 1200 E. California Blvd, Pasadena, 91125, U.S.A.
 \and
 Institute of Theoretical Astrophysics, University of Oslo, P.O. Box 1029 Blindern, N-0315 Oslo, Norway
 \and
 Centre of Mathematics for Applications, University of Oslo, P.O. Box 1053 Blindern, N-0316 Oslo, Norway
 \and
 Astrophysics Group, Cavendish Laboratory, J J Thomson Avenue, Cambridge CB3 0HE, United Kingdom
 \and
 DSM/Irfu/SPP, CEA/Saclay, F-91191 Gif-sur-Yvette C{\'e}dex, France
 \and
 Institut d'Astrophysique Spatiale, B\^atiment 121, F-91405 Orsay, France
  \and
 Institut d'Astrophysique de Paris, 98 bis Boulevard Arago, F-75014 Paris, France
 \and
 Space Sciences Laboratory, University of California Berkeley, Computational Cosmology Center,
 Lawrence Berkeley National Laboratory, CA 94720 USA
 \and
 Istituto di Scienza e Technologie dell'Informazione, CNR, Area della ricerca di Pisa, Via G. Moruzzi 1, I-56124 Pisa, Italy
}
\keywords{Methods: data analysis; Cosmology: cosmic microwave background  }
\date{September 17, 2008}

\begin{document}

\abstract{The {{\sc Planck}} satellite will map the full sky at nine frequencies
  from 30 to 857 GHz. The CMB intensity and polarization that are its prime targets
  are contaminated by foreground emission.}
{The goal of this paper is to compare proposed methods for separating
  CMB from foregrounds based on their different spectral and spatial
  characteristics, and to separate the foregrounds into `components'
  with different physical origins (Galactic synchrotron, free-free and
  dust emissions; extra-galactic and far-IR point sources;
  Sunyaev-Zeldovich effect, etc).  }
{A component separation challenge has been organised, based on a set
  of realistically complex simulations of sky emission. Several
 methods including those based on internal template subtraction, maximum 
  entropy method, parametric method, spatial and harmonic cross correlation
  methods, and independent component analysis have been tested.  
}
{ Different methods proved to be effective in cleaning the CMB
  maps of foreground contamination, in reconstructing maps of
  diffuse Galactic emissions, and in detecting point sources and
  thermal Sunyaev-Zeldovich signals. The power
  spectrum of the residuals is, on the largest scales, four orders of
  magnitude lower than the input Galaxy power spectrum at the
  foreground minimum. The CMB power spectrum was accurately recovered
  up to the sixth acoustic peak. The point source detection limit reaches 100
  mJy, and about 2300 clusters are detected via the thermal SZ effect
  on two thirds of the sky.
  We have found that no single method  performs best for all scientific objectives.}
{We foresee that the final component separation pipeline for {{\sc Planck}}
  will involve a combination of methods and iterations between
  processing steps targeted at different objectives such as diffuse
  component separation, spectral estimation, and compact source
  extraction.}
\maketitle

\section{Introduction}\label{sec:intro}

{\sc Planck} is a European Space Agency space mission whose main
objective is to measure the cosmic microwave background (CMB)
temperature and polarization anisotropies with high accuracy, high
angular resolution and with unprecedented frequency coverage
\citep{2006astro.ph..4069T}. 
In anticipation of the launch, {\sc Planck} is stimulating much
research and development into data processing methods that are capable
of addressing the ambitious science programme enabled by these
multi-frequency observations. It is expected that {\sc Planck} will
break new ground in studies of the CMB, of the interstellar medium
and Galactic emission mechanisms on scales down to a few arcminutes,
as well as of the emission from many extragalactic objects.

The processing of such multi-frequency data depends on both the
science goals, as well as on the signal to noise regime and on the
overall level and complexity of foreground contamination. This
observation is borne out by a brief historical perspective on CMB data
processing.

An example of the low foreground level and complexity regime is
provided by the observations made by Boomerang at 145, 245 and 345
GHz \citep{2006A&A...458..687M}, which targeted a region of sky with
low emission from a single dust foreground. Here the two higher
frequency channels acted as foreground monitors for the 145 GHz CMB
deep survey, and were used to estimate that the foreground
contamination at 145 GHz was at an RMS level of less than 10$\mu$K on
angular scales of $11.5'$ (Table 10, \citet{2006A&A...458..687M}).  The
145 GHz CMB maps were then used for the purpose of power spectrum
estimation in both temperature and polarization, after masking away a
handful of compact sources \citep{2006ApJ...647..823J}.
\citet{2006A&A...458..687M} estimate that the cleanest 40$\%$ of the
sky have a level of dust brightness fluctuations similar to those of
the Boomerang observations, and that the cleanest 75$\%$ of the sky
have brightness fluctuations less than three times larger.

An example of the high foreground level and complexity regime is available with the
all-sky observations of the {\sc WMAP} mission in five frequency
channels from 23 to 94 GHz \citep{2003ApJS..148....1B,2008arXiv0803.0732H}.  In this
frequency range, the emission from at least three Galactic components
(synchrotron, free-free and dust), as well as contamination by
unresolved point sources must be contended with. {\sc WMAP} also gives
a clear example of science goal dependent data processing: 
CMB maps for use in non-Gaussianity tests are obtained from a
noise-weighted sum of frequency maps at differing angular resolution,
for which the regions most contaminated by foregrounds are masked
\citep{2003ApJS..148..119K}; The analysis leading
to the {\sc WMAP} cosmological parameter estimation involves
foreground cleaning by template subtraction, masking of the most
contaminated 15$\%$ of sky, and subtracting a model of the
contribution of unresolved point sources from the CMB cross power
spectra~\citep{2003ApJS..148..135H,2007ApJS..170..288H}. For an improved
understanding of galactic emission, the WMAP team have used a number of
methods including template fits, {{\bfref internal linear combination (ILC)}},
the maximum entropy method, and the direct pixel-by-pixel fitting of
an emission model \citep{2008arXiv0803.0715G,2008arXiv0803.0586D}.

\subsection{Component Separation}

Component separation is a catch-all term encompassing any data
processing that exploits correlations in
observations made at separate frequencies, as well as external
constraints and physical modeling, as a means of distinguishing
between different physical sources of emission.

{\sc Planck} has a number of different scientific objectives: the
primary goal is a cosmological analysis of the CMB, but important
secondary goals include obtaining a better understanding of the
interstellar medium and Galactic emission, measurement of
extragalactic sources of emission and the generation of a
Sunyaev-Zeldovich (SZ) cluster catalogue.  
These planned objectives will lead to a set of data products which the
{\sc Planck} consortium is committed to delivering to the wider
community some time after the completion of the survey.
These data products include maps of the main diffuse emissions and
catalogues of extragalactic sources, such as galaxies and clusters of
galaxies.

In this context, it is worth remembering that {\sc Planck} is designed
to recover the CMB signal at the level of a few microkelvin per
resolution element of $5'$ (and less than one microkelvin per
square degree). 
Numbers to keep in mind are the RMS of CMB smoothed with a beam of
$45'$ FWHM, which is around $70 \mu$K, while the RMS of white noise at the
same scale is around $0.7 \mu$K.
This level of sensitivity sets the ultimate goals for data processing
---and component separation in particular--- if the full scientific
potential of {\sc Planck} is to be realised.
However, less stringent requirements may be acceptable for statistical
analyses such as power spectrum estimation, in particular on large
scales where cosmic variance dominates the error of total intensity observations.

\subsection{WG2}

{\sc Planck} is designed to surpass previous CMB experiments in almost
every respect.  
Therefore, a complete and timely exploitation of the
data will require methods that improve upon foreground removal via
template subtraction and masking. 
The development and assessment of such methods is coordinated within
the {\sc Planck} `Component Separation Working Group' (WG2).
Another working group in the {\sc Planck} collaboration, the $C_{\ell}$
temperature and polarization working group (WG3 or ``CTP''), investigates other
critical data analysis steps, in particular, map-making
\citep{Poutanen:2005yg,2007A&A...467..761A} and power spectrum
estimation.

The present paper reports the results of the WG2 activity in the
framework of a \emph{component separation challenge} using a common
set of simulated {\sc Planck} data.\footnote{
  A similar data challenge has been undertaken in the past in the
  context of simulated {\sc WMAP} and sub-orbital CMB data \citep[the
  WOMBAT challenge;][]{Gawiser:1998rw}.  }
 In turn, this exercise provides
 valuable feedback and validation during the development of the {\sc
Planck} Sky Model.

This is the first time, within the {\sc Planck} collaboration, that an
extensive comparison of component separation methods has been attempted on
simulated data based on models of sky emissions of representative
complexity. As will be seen and emphasised throughout this paper, this
aspect is critical for a meaningful evaluation of the performance of
any separation method.  In this respect, the present work
significantly improves on the semi-analytical estimates of foreground
contamination obtained by \citet{Bouchet:1999gq} for the {\sc Planck} phase~A study,
as well as on previous work by \citet{2000ApJ...530..133T} .

The paper is organised as follows:
In Section~\ref{sec:challenge} we describe the sky emission model and
simulations that were used, and describe the methodology of the
Challenge.  In Section~\ref{sec:spirit} we give an overview of the
methods that have been implemented and took part in the analysis.  In
Section~\ref{sec:cmb} we describe the results obtained for CMB
component separation and power spectrum estimation. The results for
point sources, SZ and Galactic components are described in
Section~\ref{sec:galcomp} and in Section~\ref{sec:summary} we present
our summary and conclusions. In Appendix~\ref{sec:methods} we provide
a more detailed description of the methods, their implementation
details and their strengths and weaknesses.

\section{The challenge}\label{sec:challenge}

The objective of the component separation challenge discussed herein
is to assess the readiness of the {\sc Planck} collaboration to tackle
component separation, based on the analysis of realistically complex
simulations.  It offers an opportunity for comparing the results from
different methods and groups, as well as to develop the expertise,
codes, organisation and infrastructure necessary for this task.

This component separation challenge is designed so as to test on
realistic simulated data sets, component separation methods and algorithms 
in a situation as close as possible to what is expected when actual {\sc Planck} data 
will be analysed.  Hence, we
assume the availability of a number of ancillary data sets.
In particular, we assumed that six-year {\sc WMAP} observations will be
available. Although {\sc WMAP} is significantly less
sensitive than {\sc Planck}, it provides very useful complementary
information for the separation of low-frequency Galactic components.
This section describes our simulations, the challenge setup, and the
evaluation methodology.

\subsection{Sky emission}
\label{sec:psm}

Our sky simulations are based on an early development version of the
{\sc Planck} Sky Model (PSM, in preparation), a flexible software package
developed by {\sc Planck} WG2 for making predictions, simulations and
constrained realisations of the microwave sky.

The CMB sky is based on the observed {\sc WMAP} multipoles up
to $\ell=70$, and on a Gaussian realisation assuming the {\sc WMAP}
best-fit $C_{\ell}$ at higher multipoles. It is the same CMB map used by 
\citet{2007A&A...467..761A}.

The Galactic interstellar emission is described by a three component model of the
interstellar medium comprising of free-free, synchrotron and dust
emissions. The predictions are based on 
a number of sky templates which have different angular resolution.
In order to simulate the sky at {\sc Planck} resolution we have added
small scale fluctuations to some of the templates. The procedure used
is the one presented in \citet{2007A&A...469..595M}
which allows to increase the fluctuation level as a function of
the local brightness and therefore reproduce the non-Gaussian properties
of the interstellar emission.

Free-free emission is based on the model of \citet{2003MNRAS.341..369D}
assuming an electronic temperature of 7000~K.
The spatial structure of the emission is
estimated using a H$\alpha$ template corrected for dust extinction.  
The H$\alpha$ map is a combination of 
the Southern H-Alpha Sky Survey Atlas (SHASSA) and the Wisconsin H-Alpha Mapper (WHAM). 
The combined map was smoothed to obtain a uniform angular resolution of 1$^{\circ}$.
For the extinction map we use the $E(B-V)$ all-sky map of
\cite{schlegel1998} which is a combination of a smoothed IRAS
100~$\mu$m map (with resolution of $6.1'$) and a map at a few degrees
resolution made from DIRBE data to estimate dust temperature and 
transform the infrared emission in extinction. As mentioned earlier, small
scales were added in both templates to match the {\sc Planck} resolution.
\begin{figure}[]
  \centering
  \resizebox{\hsize}{!}{\includegraphics[angle=90, draft=\draftfig]{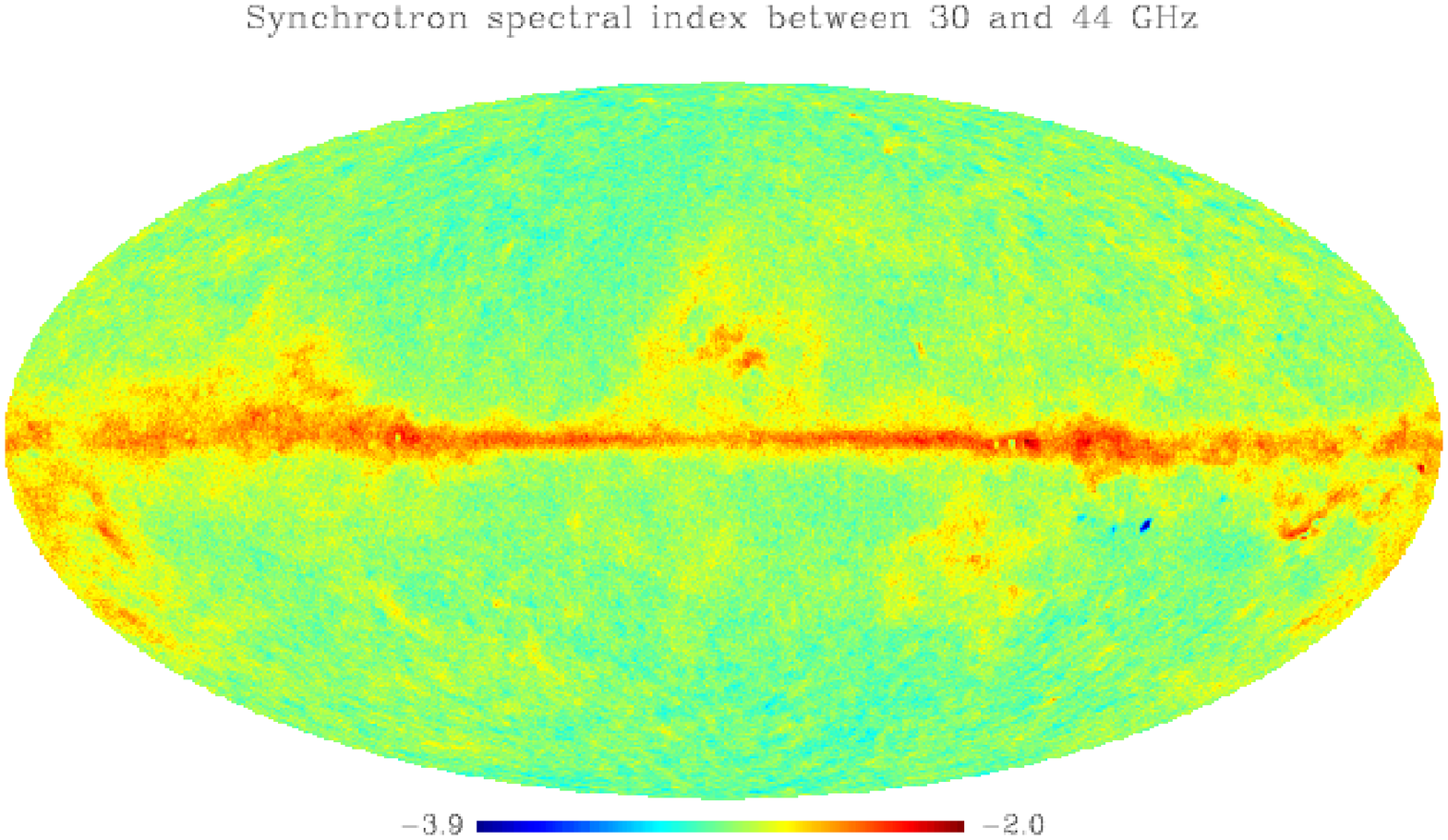}}
  \resizebox{\hsize}{!}{\includegraphics[angle=90, draft=\draftfig]{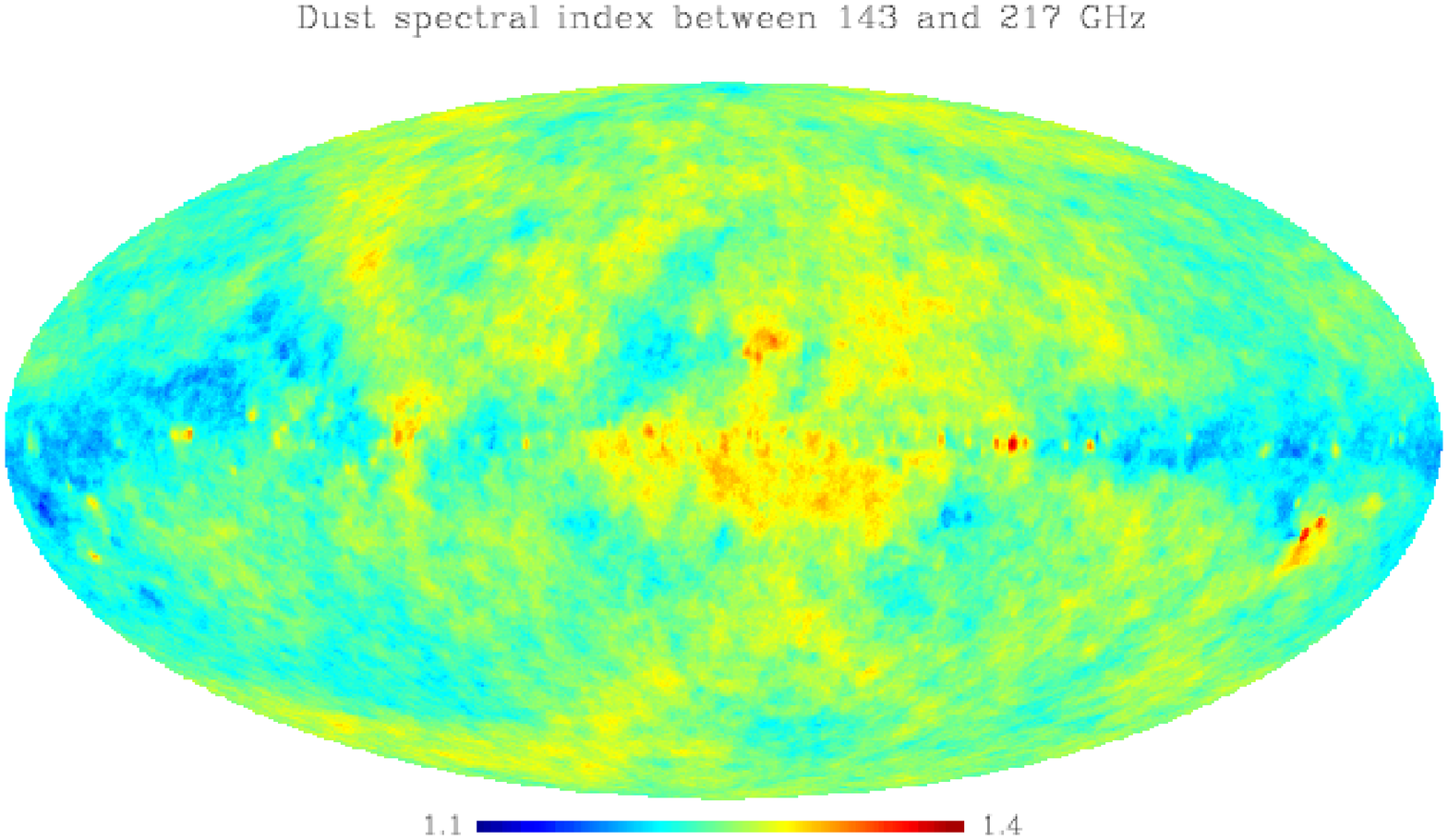}}
  \caption{Synchrotron and dust (effective powerlaw) spectral indices evaluated between 30
    and 44 GHz, and 143 and 217 GHz respectively (in $\mu$K$_{\rm RJ}$).  A spatially varying
    spectral index corresponds to a foreground morphology that varies
    with frequency.}
  \label{fig:specindex}
\end{figure}

Synchrotron emission is based on an extrapolation of the 408~MHz map of \cite{haslam1982}
from which an estimate of the free-free emission was removed.
The spectral emission law of the synchrotron is assumed to follow a perfect
power law, $ T_b^{sync} \propto\nu^\beta$.  
We use a pixel-dependent spectral index $\beta$ derived from the ratio of the
408~MHz map and the estimate of the synchrotron emission at 23~GHz in the {\sc WMAP} data
obtained by \citet{2003ApJS..148...97B} using a Maximum Entropy Method technique. 
A limitation of this approach is that this synchrotron model 
also contains any `anomalous'  dust correlated emission seen by {\sc WMAP} at 23 GHz.

The thermal emission from interstellar dust is estimated using model
7 of \citet{finkbeiner1999}. This model, fitted to the
FIRAS data (7$^{\circ}$ resolution), makes the hypothesis that each line
of sight can be modelled by the sum of the emission from two dust
populations, one cold and one hot. Each grain population is in thermal
equilibrium with the radiation field and thus has a grey-body
spectrum, so that the total dust emission is modelled as
\begin{eqnarray}
\label{eq:dustmodel}
I_\nu \propto \sum_{i=1}^2 f_i \nu^{\beta_i} B_\nu(T_i)
\end{eqnarray}
where $B_\nu(T_i)$ is the Planck function at temperature $T_i$.  In
model 7 the emissivity indices are $\beta_1=1.5$, $\beta_2=2.6$, and $f_1=0.0309$ and $f_2 =
0.9691$.  Once these values are fixed, the dust temperature of the two
components is determined using only the ratio of the observations at
two wavelengths, 100~$\mu$m and 240~$\mu$m.  For this purpose, we use
the 100/240~$\mu$m map ratio published by \cite{finkbeiner1999}.
Knowing the temperature and $\beta$ of each dust component at a given
position on the sky, we use the 100~$\mu$m brightness at that position
to scale the emission at any frequency using Eq.(~\ref{eq:dustmodel}).
We emphasise that the emission laws of the latter two components,
synchrotron and dust, vary across the sky as shown in
Figure~\ref{fig:specindex}. The spectral index of free-free is taken to be
uniform on the sky since it only depends on the electronic temperature,
taken as a constant here.

Point sources are modelled with two main categories: radio and
infra-red.
Simulated radio sources are based on the NVSS or SUMSS and GB6 or PMN
catalogues. Measured fluxes at 1 and/or 4.85 GHz are extrapolated to
{\sc Planck} frequencies assuming a distribution in flat and steep
populations. For each of these two populations, the spectral index is
randomly drawn from within a set of values compatible with the typically observed
mean and dispersion.
Infrared sources are based on the {\sc IRAS} catalogue, and modelled as
dusty galaxies~\citep{2005MNRAS.356..192S}. IRAS coverage gaps 
were filled by adding simulated sources with a flux distribution consistent with 
the mean counts.  Fainter sources were assumed to be mostly sub-millimeter 
bright galaxies, such as those detected by SCUBA surveys. These were 
modelled following \citet{2004ApJ...600..580G} and assumed to be strongly 
clustered, with a comoving clustering radius $r_0 \simeq 
8\,\hbox{h}^{-1}\,$Mpc. Since such sources have a very high areal density, 
they are not simulated individually but make up the sub-mm background.

We also include in the model a map of thermal SZ spectral distortions
from galaxy clusters, based on a simulated cluster catalogue 
drawn from a mass-function compatible with present-day observations and with
$\Lambda$CDM parameters $\Omega_{\rm m}=0.3$, $h=0.7$ and $\sigma_8=0.9$
\citep{1997ApJ...479....1C,2005A&A...431..893D}.

Component maps are produced at all {\sc Planck} and \textsc{WMAP} central
frequencies.  They are then co-added and smoothed with Gaussian beams
as indicated in Table~\ref{tab:freqchann_Planck}.  A total of fourteen
monochromatic maps have been simulated.

\begin{figure}[t]
  \centering
  \resizebox{\hsize}{!}{\includegraphics[angle=90, draft=\draftfig]{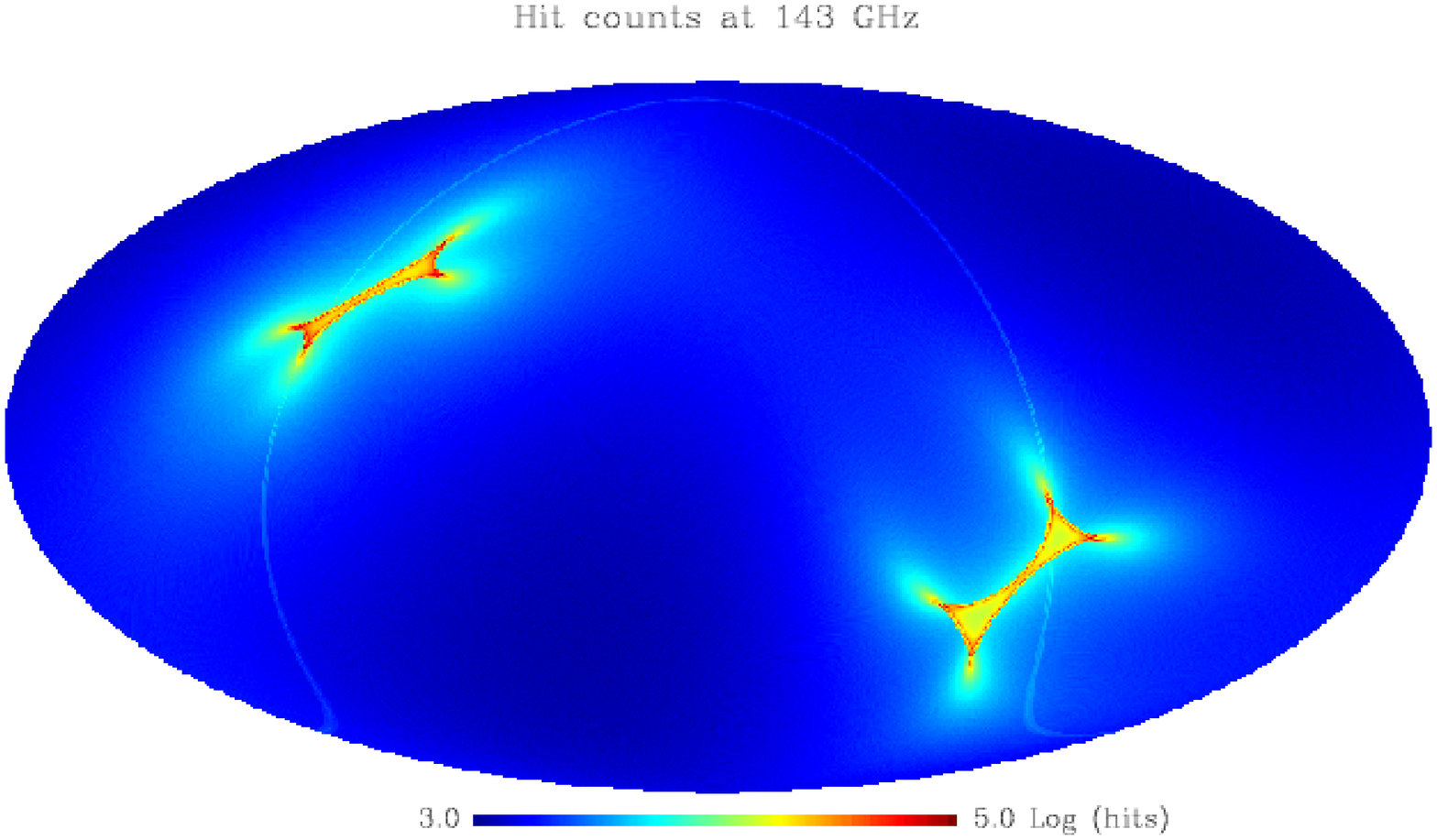}}
  \resizebox{\hsize}{!}{\includegraphics[angle=90, draft=\draftfig]{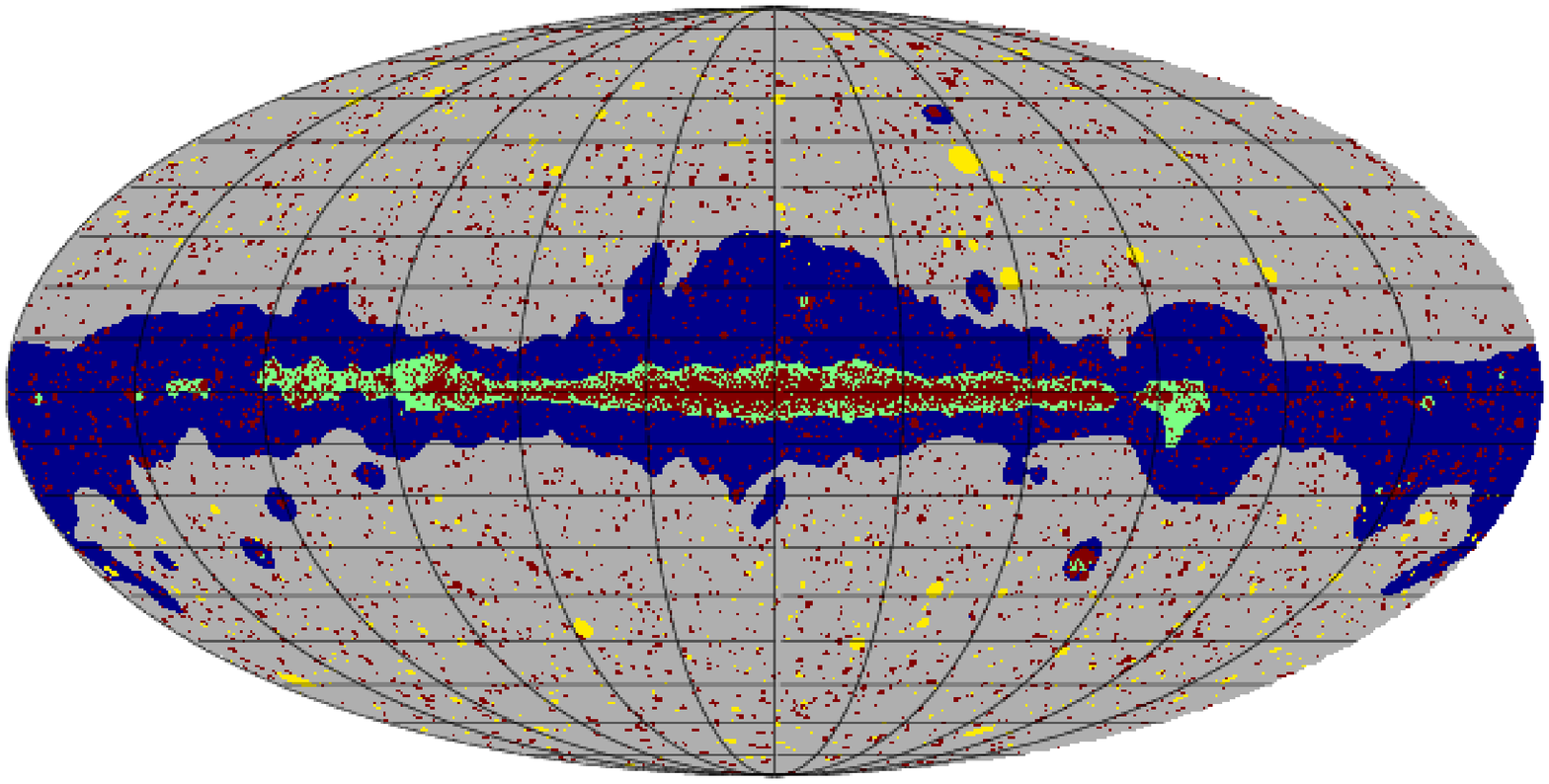}}
  \caption{Upper panel: Hit counts for the 143~GHz channel. The
    inhomogeneities at the ecliptic poles are characteristic of {\sc
      Planck}'s cycloidal scanning strategy.  
    Lower panel: The masking scheme separating the sky in three
    regions of different foreground contamination.  The grey region at
    high Galactic latitudes is Zone~1, covering $f_{{\rm
        sky}}=74\%$. The darker region at lower Galactic latitudes is
    Zone~2 and covers $f_{{\rm sky}}=22\%$. The remaining region (green) 
    along the Galactic ridge is Zone~3. The point source mask (red)
    covers $4\%$ of Zone~1. The SZ mask (yellow) cuts detected SZ clusters
    at Galactic latitudes above 20 degrees, covering $1.4\%$ of sky.
}
  \label{fig:hitmap}
\end{figure}

Finally, inhomogeneous noise is obtained by simulating the hit counts
corresponding to one year of continuous observations by {\sc Planck},
using the Level-S simulations tool \citep{2006A&A...445..373R}. An
example of a hit count map is shown in the upper panel of
Figure~\ref{fig:hitmap}.  {\sc WMAP} six year hit counts, obtained
from scaling up the observed {\sc WMAP} three year hit count patterns,
are used to generate inhomogeneous noise in the simulated {\sc WMAP}
observations.  The RMS noise level per hit for both experiments is
given in Table~\ref{tab:freqchann_Planck}.

\begin{table*} 
  \caption{{\bf Characteristics of {\sc Planck} one year simulations (upper) and {\sc WMAP} six year simulations (lower)}.
    {\sc Planck} and {\sc WMAP} hit counts correspond to $1.7'$ (Healpix $n_{\rm side}$=2048) and
    $6.8'$ ($n_{\rm side}$=512) pixels respectively. $N_{\ell}$ is the white noise level calculated from
    the inhomogeneous distribution of hits. }
  \centering
  \begin{tabular}{c|ccccccccc}
    \hline
    \hline
    Channel & 30 GHz & 44 GHz & 70 GHz & 100 GHz & 143 GHz & 217 GHz & 353 GHz & 545 GHz & 857 GHz \\ \hline
    FWHM [arcmin]  & 33 & 24 & 14 &  10 & 7.1 & 5 & 5 & 5 & 5 \\ 
    $\sigma_{\rm hit}$ [$\mu$K$_{\rm RJ}$] & 1030 & 1430. & 2380 &  1250 & 754 & 610 & 425& 155 & 72 \\
    $\sigma_{\rm hit}$ [$\mu$K$_{\rm CMB}$] & 1050 & 1510 & 2700  & 1600 & 1250 & 1820 & 5470  & 24700 & 1130000  \\
    Mean; Median hits per pixel & 82; 64 & 170; 134 & 579; 455 & 1010; 790  & 2260; 1790  & 2010; 1580  & 2010; 1580 & 503; 396 & 503; 396  \\ 
    $N_{\ell}^{1/2}$ [$\mu$K$_{\rm CMB}$] & 0.066 & 0.065 & 0.063 & 0.028  & 0.015 & 0.023 & 0.068 &0.62  & 28.4   \\ \hline
  \end{tabular}
  \vspace{0.3 cm}
  \\
  \begin{tabular}{c|ccccccccc}
    \hline
    \hline
     Channel & 23 GHz (K) & 33 GHz (Ka) & 41 GHz (Q) & 61 GHz (V) & 94 GHz (W) \\ \hline
    FWHM [arcmin] & 52.8 & 39.6 & 30.6 & 21 & 13.2 \\
    $\sigma_{\rm hit}$ [$\mu$K$_{\rm RJ}$] &  1420  & 1420 & 2100 &  2840 & 5210 \\
    $\sigma_{\rm hit}$ [$\mu$K$_{\rm CMB}$] & 1440 & 1460 & 2190 & 3120 & 6500 \\
    Mean; Median hits per pixel &  878; 792 & 878; 790 & 2198; 1889 & 2956; 2577 & 8873; 7714 & \\ 
    $N_{\ell}^{1/2}$ [$\mu$K$_{\rm CMB}$] & 0.10  & 0.10  & 0.10  & 0.12 & 0.14  & \\ \hline
  \end{tabular}
  \label{tab:freqchann_Planck}
\end{table*}

\subsection{Challenge setup}

The simulated data sets were 
complemented by a set of ancillary data including hitmaps and noise
levels, IRAS, 408 MHz, and H$\alpha$ templates, as well as catalogues
of known clusters from ROSAT and of known point sources from NVSS,
SUMSS, GB6, PMN and IRAS.

The Challenge proceeded first with a blind phase lasting around
four months between August and November 2006, when neither the exact
prescription used to simulate sky emission from these ancillary data
sets, nor maps of each of the input components, were communicated to
challenge participants.

After this phase and
an initial review of the results at the WG2 meeting in Catania in
January 2007, the Challenge moved to an open phase lasting from January
to June 2007. In this phase the input data---CMB maps and power
spectrum, Galactic emission maps, SZ Compton $y$ parameter map, point source
catalogues and maps, noise realisations---were made available to the
participating groups.

All of the results presented here have been obtained after several
iterations and improvements of the methods, both during the comparison
of the results obtained independently by the various teams, and after
the input data was disclosed. Hence, the challenge has permitted
significant improvement of most of the methods and algorithms
developed within the {\sc Planck} collaboration. The analysis of the
Challenge results was led by the simulations team, with involvement
and discussion from all participating groups.

\subsubsection*{Deliverables}

A set of standard deliverables were defined. These included: 
a CMB map with $1.7'$ pixels (Healpix $N_\mathrm{side}=2048$) 
together with a corresponding map of estimated errors;
the effective beam $F_{\ell}$, which describes the total
smoothing of the recovered CMB map due to a combination of
instrumental beams and the filtering induced by the component
separation process;
a set of binned CMB power spectrum estimates (band averages of
$\ell(\ell + 1)C_{\ell}$) and error bars; 
maps of all the diffuse components identified in the data; 
catalogues of the infrared and radio sources, and SZ clusters;
a map of the SZ Compton $y$ parameter.

\subsubsection*{Masks}

Different separation methods are likely to perform differently in
either foreground-dominated or noise-dominated observations.  Also,
they may be more or less sensitive to different types of foregrounds.
Since the level of foreground contamination varies strongly across the
sky, we used a set of standard masks throughout this work, and  they are shown
in the lower panel of Figure~\ref{fig:hitmap}.

The sky is split into three distinct Galactic `Zones': Zone~1 is at high
Galactic latitudes and covers $74\%$ of sky, similar to the WMAP Kp0
mask with smoother edges and small extensions.  Zone~2 is at lower
Galactic latitudes and covers $22\%$ of sky. The remaining $4\%$
of sky is covered by Zone~3, which is similar to the WMAP Kp12 mask.

The point source mask is the product of nine masks, each constructed
by excluding a two FWHM region around every
source with a flux greater than 200~mJy at the corresponding {\sc Planck}
frequency channel.
This point source mask covers $4\%$ of sky in Zone~1.
For comparison, the {\sc WMAP} point source masks of
\citet{2003ApJS..148...97B} excludes a radius of 0.7$^{\circ}$ around
almost 700 sources with fluxes greater than 500~mJy, covering a total
of 2$\%$ of sky.

The SZ mask is constructed by blanking out small circular regions
centered on 1625 SZ clusters detected with the needlet-ILC + matched filter method (see
Section~\ref{sec:szresults}).  For each of them, the diameter of the
cut is equal to the virial radius of the corresponding cluster.

\subsection{Comments about the sky emission simulations}
\label{sec:comments}

A note of caution about these simulations of sky emission is in order.  
Although the PSM, as described
above, has a considerable amount of sophistication, it still makes some
simplifying assumptions -- and cannot be expected to describe the full
complexity of the real sky.  This is a critical issue, as component
separation methods are very sensitive to these details.  We mention
four of them.

First, Galactic emission is modelled with only three components, with
no anomalous emission at low frequencies. This affects the spectral 
behaviour of components in the lower frequency bands below 60~GHz 
where the anomalous emission is thought to be
dominant \citep{2006MNRAS.370.1125D,2007MNRAS.382.1791B,2008arXiv0802.3345M}.

Second, even though variable spectral emission laws are used for
synchrotron and dust emission, this is still an idealisation: for the
synchrotron, the emission law in each pixel is described by a single
spectral index without any steepening.
For dust, the emission is modelled as a superposition of two populations,
with distinct but fixed temperature and emissivity.
These approximations impact component separation, since almost perfect
estimation of the relevant parameters of a given foreground emission
is possible at frequencies where this foreground dominates, thereby
allowing perfect subtraction in the cosmological channels.

Third, it is worth mentioning that only low resolution ($\sim 1^\circ$) templates are
available for synchrotron and free-free emissions.  Hence, addition of
small-scale power is critical: if such scales were absent from the
simulations, but actually significant in the real sky, one might get
a false impression that no component separation is needed on small
scales.  Also, the detection of point sources as well as galaxy
clusters would be significantly easier, hence not representative of
the actual problem.
Here, missing small scale features are simulated using a
non-stationary coloured Gaussian random field. Although quite
sophisticated, this process can not generate for instance, filamentary
or patchy structures known to exist in the real sky.

\begin{figure}[t]
  \centering
  \resizebox{\hsize}{!}{\includegraphics[draft=\draftfig]{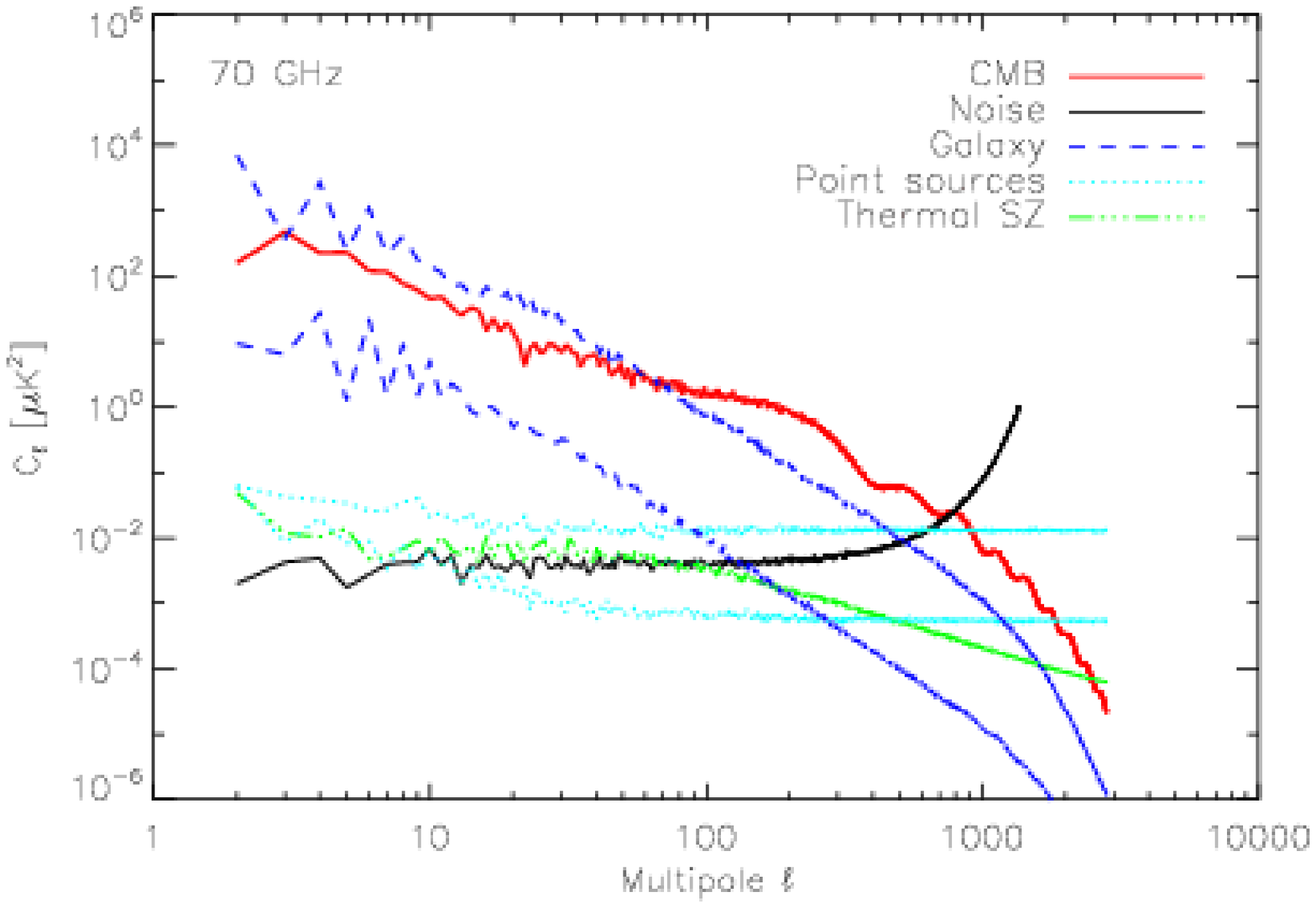}}
  \resizebox{\hsize}{!}{\includegraphics[draft=\draftfig]{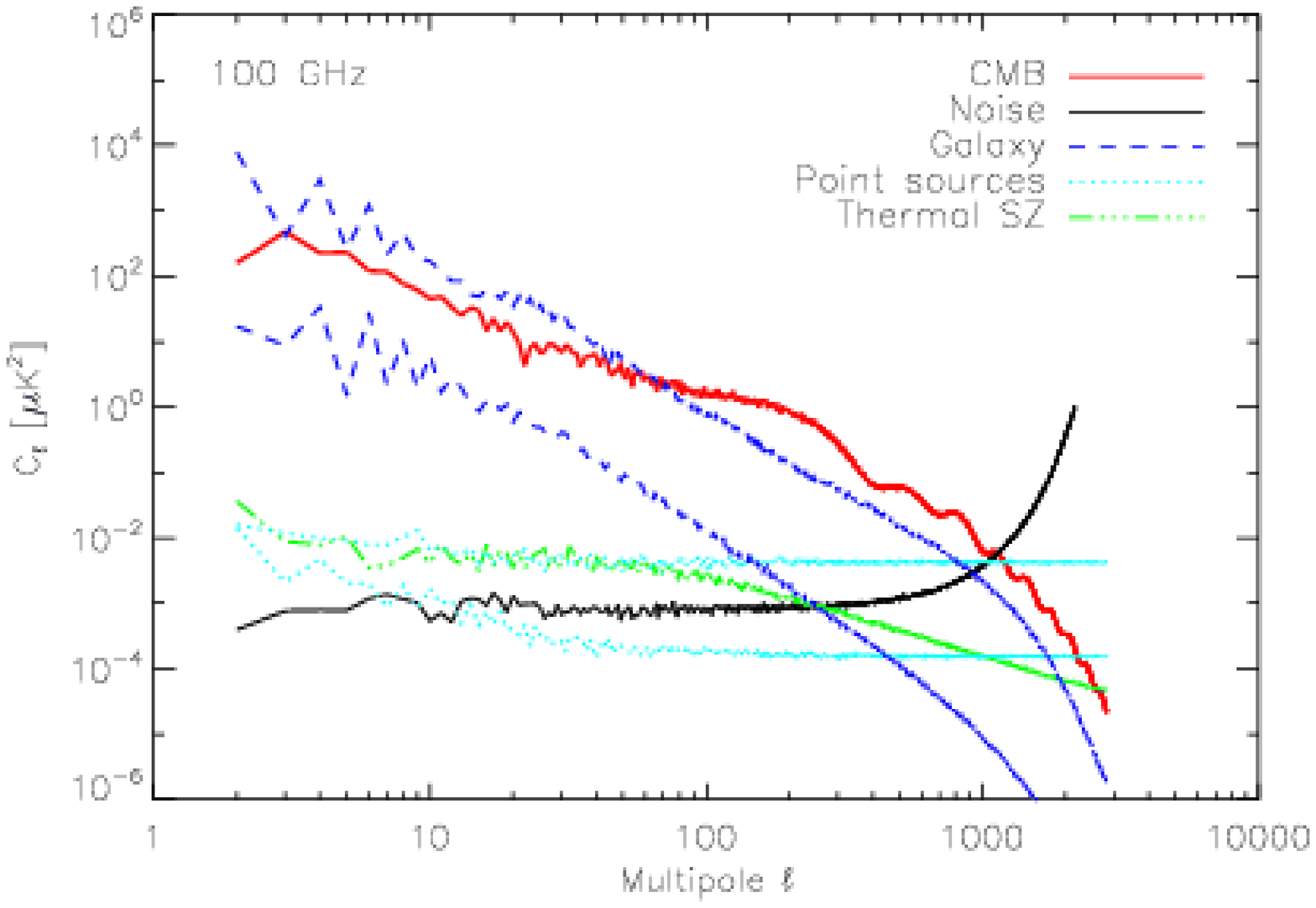}}
  \caption{Spectra of the simulated microwave sky components near the
    foreground minimum. CMB, noise with the effect of beam
    deconvolution, and the thermal SZ effect are evaluated on the full
    sky; point source power is evaluated on Zone~1+2 both with and without
    sources above 200mJy masked; the galaxy power spectra are
    evaluated on Zone~1 and on Zone~2.  The well-known importance of
    masking is evident, as is the fact that there is a significant
    proportion of sky (Zone~2, $f_{{\rm sky}}=22\%$) for which
    Galactic emission is comparable to CMB power.}
  \label{fig:clspectra}
\end{figure}

Fourth, our simulations are somewhat idealised in the sense that we use
perfect Gaussian beams, assume no systematic effects, and assume that
the noise is uncorrelated from pixel to pixel and from channel to
channel.  Also, the effect of the finite bandpass of the frequency
channels is not taken into account, and we assume that the calibration
and zero levels of each channel is perfectly known.

In spite of these simplifications, component separation remains a
difficult task with our simulated data because of pixel-dependent
spectral emission laws for dust and synchrotron, and of the presence of
more than a million point sources with different emission laws, of
hundreds of thousands of unresolved or extended SZ clusters, and of
significant emission from a complex IR background.
It is fair to say that this simulated sky is far more complex than anything ever
used in similar investigations.

In closing this Section, we show in Figure~\ref{fig:clspectra}
the angular power spectra of
the basic components for the 70 and 100 GHz channels, close to the
foreground emission minimum. The spectra of CMB, noise and thermal SZ
are compared to the spectra of the total Galactic emission evaluated at high and low
Galactic latitudes, on Zone~1 and 2 respectively.  
The point source spectra are evaluated in Zone~1, both with and
without the brightest sources above 200~mJy masked.
Figure~\ref{fig:clspectra} shows the obvious impact on CMB studies of
masking the most foreground-contaminated regions.  It also indicates
that there is a significant region of sky, Zone~2, for which Galactic
emission and CMB power are comparable. In the following sections,
results are evaluated independently in both Zones 1 and 2.

\section{Outline of the methods}\label{sec:spirit}

In this section we present a brief overview of the methods that have
been used in this challenge. The section is divided in three parts,
one for diffuse component separation methods, one for point source
extraction, and one for SZ cluster extraction.

\subsection{Diffuse component separation}

The spirit of each
method tested on the challenge data is outlined here. A more detailed
description, including some details of their implementations and a
discussion of their strengths and weaknesses is presented in
Appendix~\ref{sec:methods}.

\begin{table*}
  \centering
  \caption{{\bf Some characteristics of the diffuse component separation methods used in the challenge.}}
  \begin{tabular}{c|cccc}
    \hline
    \hline
                      &Channels used      & Components modelled &   Resources and runtime \\
    \hline  COMMANDER & {\sc WMAP}, {\sc Planck} 30--353 GHz,      & CMB, dust, sync, FF, mono-,dipoles  & 1000 CPU hr, 2 day\\
    \hline  CCA     & {\sc Planck}, Haslam 408 MHz       & CMB, dust, sync, FF & 70 CPU hr, 1.5 day\\
    \hline  GMCA    & {\sc Planck}, Haslam 408 MHz       & CMB, SZ, sync., FF       & 1200 CPU hr, 6 day \\
    \hline  FastICA & 143--353 GHz         & Two components (CMB and dust)          & 21 CPU min, 20 sec\\
    \hline  FastMEM  & {\sc Planck}         & CMB, SZ, dust, sync, FF  & 256 CPU hr, 8 hr\\
    \hline  SEVEM  & {\sc Planck}        & CMB   & 30 CPU hr, 30 hr\\
    \hline  SMICA  & {\sc Planck}, {\sc WMAP}  & CMB, SZ, dust, total galaxy & 8 CPU hr, 4 hr\\
    \hline  WI-FIT & 70--217 GHz     & CMB        & 400 CPU hr, 8 hr\\
    \hline
  \end{tabular}
  \label{tab:methods}
\end{table*}

First we define some relevant terminology. The data model for a given channel $\nu$ is
\begin{eqnarray}
  \label{eq:obsmod}
  d_\nu  = b_\nu * x_\nu + n_\nu
\end{eqnarray}
where $d_\nu$, $x_\nu$, $n_\nu$ are respectively the observation map,
the sky emission map and the noise map at frequency $\nu$ while
$b_\nu$ is the instrumental beam of channel $\nu$, assumed to be
Gaussian symmetric, and $*$ denotes convolution on the sphere.
The sky emission itself, $x_\nu$, is a superposition of components.
Most methods assume (implicitly or explicitly), that it can be written
as a linear mixture
\begin{eqnarray}
  \label{eq:skymixing}
  x_\nu = \sum_c A_{\nu c} s_c
\end{eqnarray}
where the sum runs over the components. 
In matrix-vector format, this reads $\Bx=\BA\Bs$ where $\BA$ is
referred to as the `mixing matrix'.  Vector $\Bs$ is the vector of
components.  Vectors $\Bd$ and $\Bn$ are defined similarly.
When this model holds, Eq.~(\ref{eq:obsmod}) becomes
\begin{eqnarray}
  \label{eq:mixing}
  d_\nu  = b_\nu * \left( \sum_c A_{\nu c} s_c \right)  + n_\nu
\end{eqnarray}
In simple models, matrix $\BA$ is constant over the sky; in more
complex models, it varies over patches or even from pixel to pixel.

We now briefly describe each of the methods that performed component separation of 
the CMB (and possibly other diffuse components), and also mention how the
CMB angular power spectrum is estimated.

\begin{itemize}
\item \textbf{Gibbs sampling} \citep[Commander;][]{2008ApJ...676...10E}.
  The approach of Commander is to fit directly an explicit parametric
  model of CMB, foregrounds and noise to the antenna temperature of
  low-resolution map pixels. For the Challenge, Commander was used to analyse
  the data smoothed to $3^{\circ}$ resolution at each channel with a pixel
  size of $54'$ (Healpix $N_{\rm side}$=64).
  For a given foreground model, Commander provides an exact
  foreground-marginalised CMB $C_{\ell}$ distributions using the Gibbs
  (conditional) sampling approach. 
\item \textbf{Correlated component analysis} \citep[CCA;][]{ccapisa}.
  The CCA approach starts with an estimation of the mixing matrix on
  patches of sky by exploiting spatial correlations in the data,
  supplemented by constraints from external templates and foreground
  scaling modeling.  The estimated parameters are then used to reconstruct the
  components by Wiener filtering in the harmonic domain. The $C_\ell$ are
  estimated from the recovered CMB map.
\item \textbf{Independent component analysis}
  \citep[FastICA;][]{2002MNRAS.334...53M}.  
  The FastICA method is a popular approach to blind component
  separation.  
  No assumptions are made about the frequency scaling or mixing matrix.
  Instead, assuming statistical independence between CMB and
  foregrounds, the mixing matrix is estimated by maximizing the
  non-gaussianity of the 1-point distribution function of linear
  combinations of input data.
  The inferred mixing matrix is used to invert the linear system of
  Eq.~(\ref{eq:mixing}). The $C_\ell$'s are estimated from the recovered CMB map.
\item \textbf{Harmonic-space maximum entropy method} \citep[FastMEM;][]{1998MNRAS.300....1H,mem}.
  The FastMEM method estimates component maps  
  given frequency scaling models and external foreground power spectra  
  (and cross-power spectra) with adjustable prior weight. It is a non- 
  blind, non-linear approach to inverting Eq.~(\ref{eq:mixing}),
  which assumes a maximum-entropy prior probability distribution for
  the underlying components.  
  The $C_{\ell}$'s are estimated from the recovered CMB component.
\item \textbf{Generalised morphological component analysis}
  \citep[GMCA;][]{gmca}.  Generalised Morphological Component Analysis
  is a semi-blind source separation method which disentangles the
  components by assuming that each of them is sparse in a fixed
  appropriate waveform dictionary such as wavelets. For the Challenge
  two variants of GMCA were applied: GMCA-blind was optimised for separation
  of the CMB component, and GMCA-model was optimised for separation of
  galactic components.  
  The $C_\ell$'s are estimated from the recovered CMB map from the
  GMCA-blind method.
\item \textbf{Spectral estimation via expectation maximisation}
  \citep[SEVEM;][]{Martinez-Gonzalez:2003}.  
  SEVEM performs component separation in three steps. In a first step,
  an internal template subtraction is performed
  in order to obtain foreground-reduced CMB maps in three centre
  channels (100-217~GHz).  Then the CMB power spectrum is estimated
  from these maps, via the EM algorithm, assuming a signal plus
  (correlated) noise model.  A final CMB map is obtained using a
  harmonic Wiener filter on the foreground-reduced maps.
\item \textbf{Spectral matching independent component analysis}
  \citep[SMICA;][]{2003MNRAS.346.1089D,2008arXiv0803.1814C}.  
  The SMICA method estimates model parameters using observation
  correlations in the harmonic domain (auto- and cross-spectra).  The
  estimated parameters are typically some mixing coefficients and the power
  spectra of independent components. For the challenge,
  the correlations between Galactic components are taken into account.  The
  estimated parameters are then used to Wiener-filter the observations
  to obtain component maps. At small scales the $C_\ell$'s are  one of the
  estimated parameters. At large scales  $\ell\leq 100$ the $C_{\ell}$'s
  are estimated from a CMB map {{\bfref obtained using the ILC method}}.
\item \textbf{Wavelet based high resolution fitting of internal
  templates} \citep[WI-FIT;][]{Hansen:2006rj}. The WI-FIT method
  computes CMB-free foreground plus noise templates from differences of the observations 
  in different channels, and uses those to fit
  and subtract foregrounds from the CMB dominated channels in wavelet space.
  The $C_\ell$'s are estimated from the recovered CMB map.
\end{itemize}
Some characteristics of these methods are summarised in
Table~\ref{tab:methods}, which shows the data used, the components
modelled and a rough indication of the computational resources required.

Note that many different approaches to diffuse component separation
are represented here: blind, non-blind, semi-blind; methods based on
linear combinations for foreground extraction; likelihood based
methods which estimate parameters of a model of the foregrounds and
the CMB; a maximum entropy method; methods based on cross
correlations; a method based on sparsity.  They also rely on very
diverse assumptions and models.

\subsection{Point source extraction}
\label{sec:psextr}

In the present challenge, point sources are detected in all {\sc
  Planck} channels independently.  Two methods are used, the first
based on a new implementation of matched filtering, and the second
using the second member of the Mexican Hat Wavelet Family of filters
\citep{gnuevo06}.  Point sources are detected by thresholding on the
filtered maps.

This corresponds to a first step for effective point source detection.
It does not exploit any prior information on the position of candidate
sources;  Such information can be obtained from external catalogues as
in \cite{2007ApJS..170..108L}, or from detections in other {\sc Planck} channels. Neither
does this approach exploit the coherence of the contaminants
throughout {\sc Planck} frequencies, nor try to detect point sources
\emph{jointly} in more than one channel.  Hence, there is margin for
improvement.

\begin{itemize}
\item {\bf Matched Filter (MF)}: 
  The high spatial variability of noise and foreground emission
  suggests using local filters (for instance on small patches).  The
  sky is divided into 496 overlapping circular regions 12 degrees in
  diameter.  Matched filtering is applied on each patch independently.
  A local estimate of the power spectrum of the background is obtained
  from the data themselves by averaging the power in circular
  frequency bins.  A first pass is performed to detect and remove the
  brightest sources (above 20$\sigma$), in order to reduce the bias in
  background power estimation and to reduce possible artifacts in the
  filtered maps.  Having removed these bright sources, the 5$\sigma$
  level catalogue is obtained by a second application of the whole procedure.

\item {\bf Mexican Hat Wavelet (MHW2)}: In a similar way, the sky is
  divided into 371 square patches. The size of each patch is
  $14.65\times 14.65$ square degrees, with a 3 degree overlap among
  patches. Each patch is then individually filtered with the
  MHW2. For each patch, the optimal scale of the wavelet is obtained
  by means of a fast maximization of the wavelet gain factor. This
  step requires only a straightforward estimation of the variance of
  the patch, excluding the border and masking any sources 
  above 30$\sigma$. A 5$\sigma$ level
  catalogue is obtained by simple thresholding in a single step.

\end{itemize}

\subsection{SZ cluster extraction}
\label{sec:szextr}

In the present data challenge, we address both the question of
building an SZ catalogue, and of making a map of thermal SZ emission.

\smallskip\noindent {\bf SZ map:} Three methods successfully produced
SZ maps: ILC in harmonic space, ILC on a needlet frame, and SMICA.
For ILC methods, the data are modelled as $\vec{d} = \vec{a} s +
\vec{n}$ where $\vec d$ is the vector of observations (nine maps here,
using {\sc Planck} data only), $\vec a$ is the SZ spectral signature at
all frequencies (a vector with nine entries), {{\bfref $s$ is the component
amplitude}} and $\vec n$ is the
noise.  The ILC provides an estimator $\widehat{s}_{\rm ILC}$ of $s$
using
\begin{eqnarray}
  \widehat{s}_{\rm ILC} = \frac{\vec{a}^t \, {\widehat{\tens{R}}}^{-1}}{\vec{a}^t \, {\widehat{\tens{R}}}^{-1} \, \vec{a}} \, \vec{d}
  \label{eq:ILC}
\end{eqnarray}
where ${\widehat{\tens{R}}}$ is the empirical correlation of the
observations, i.e. a $9 \times 9$ matrix, with entries $R_{\nu\nu'}$.
In practice, the filter is implemented in bands of $\ell$ (ILC in
harmonic space) or on subsets of needlet coefficients (ILC in needlet
space).  The needlet-ILC adapts to the local background to recover the
SZ sky.

\smallskip\noindent {\bf SZ catalogue:} Three main methods were used to obtain the
cluster catalogue:
\begin{itemize}

\item The first one uses a single frequency matched filter \citep{2006A&A...459..341M}
  to extract clusters from the needlet-ILC map.
  
\item The second one uses SExtractor~\citep{1996A&AS..117..393B} to
extract clusters from the needlet-ILC map. Then, a single frequency
matched filter is used to estimate cluster fluxes.

\item The third is a Matched MultiFilter~\citep{2002MNRAS.336.1057H},
  which implements cluster detection using the full set of input
  observations rather than from an intermediate SZ map. This third
  method is implemented independently in Saclay and in Santander.

\end{itemize}
The performance of these four methods is detailed  in
Table~\ref{tab:sz_performances}.  The comparison is done at the same
contamination level ($\sim$~10\%), which corresponds to $S/N>4.7$ for
the needlet-ILC + MF catalogue, $S/N>3.8$ for the needlet-ILC +
SExtractor catalogue, $S/N>4.3$ for the Matched Multifilter (MMF)
Saclay catalogue and $S/N>4.6$ for the MMF Santander catalogue.

This comparison is being extended to other cluster extraction methods
in collaboration with the {\sc Planck} `Clusters and Secondary
Anisotropies' working group (WG5). Some improvements are obtained
using SExtractor as the extraction tool after the component separation
step. There is still some margin for other improvements by increasing
the studied area to include lower Galactic latitudes and by combining the SZ
extraction methods with CMB and Galactic extraction methods more
intimately.

\section{Results for CMB}\label{sec:cmb}

We now turn to the presentation and discussion of the results of the
challenge, starting with the CMB component.  We evaluate performance
based on residual errors at the map and spectral level, and on
residual errors at the power spectrum estimation level.

The first point to be made is that all methods have produced CMB maps
in Zones 1 and 2.  Foreground contamination is barely visible.
A small patch representative of CMB reconstruction at
intermediate Galactic latitude, is shown in
Figure~\ref{fig:comp_maps}.  In the following, we focus on the
analysis of the reconstruction error (or residual).

Since each method produces a CMB map at a different resolution,
the recovered CMB maps are compared both against the input CMB sky
smoothed only by the $1.7'$ pixels, and against a $45'$ smoothed version,
in order to the emphasise errors at large scales.

\subsection{Map-level residual errors}
\label{sec:maplevel}

\begin{figure}[t]
  \centering
  \resizebox{\hsize}{!}{\includegraphics[draft=\draftfig]{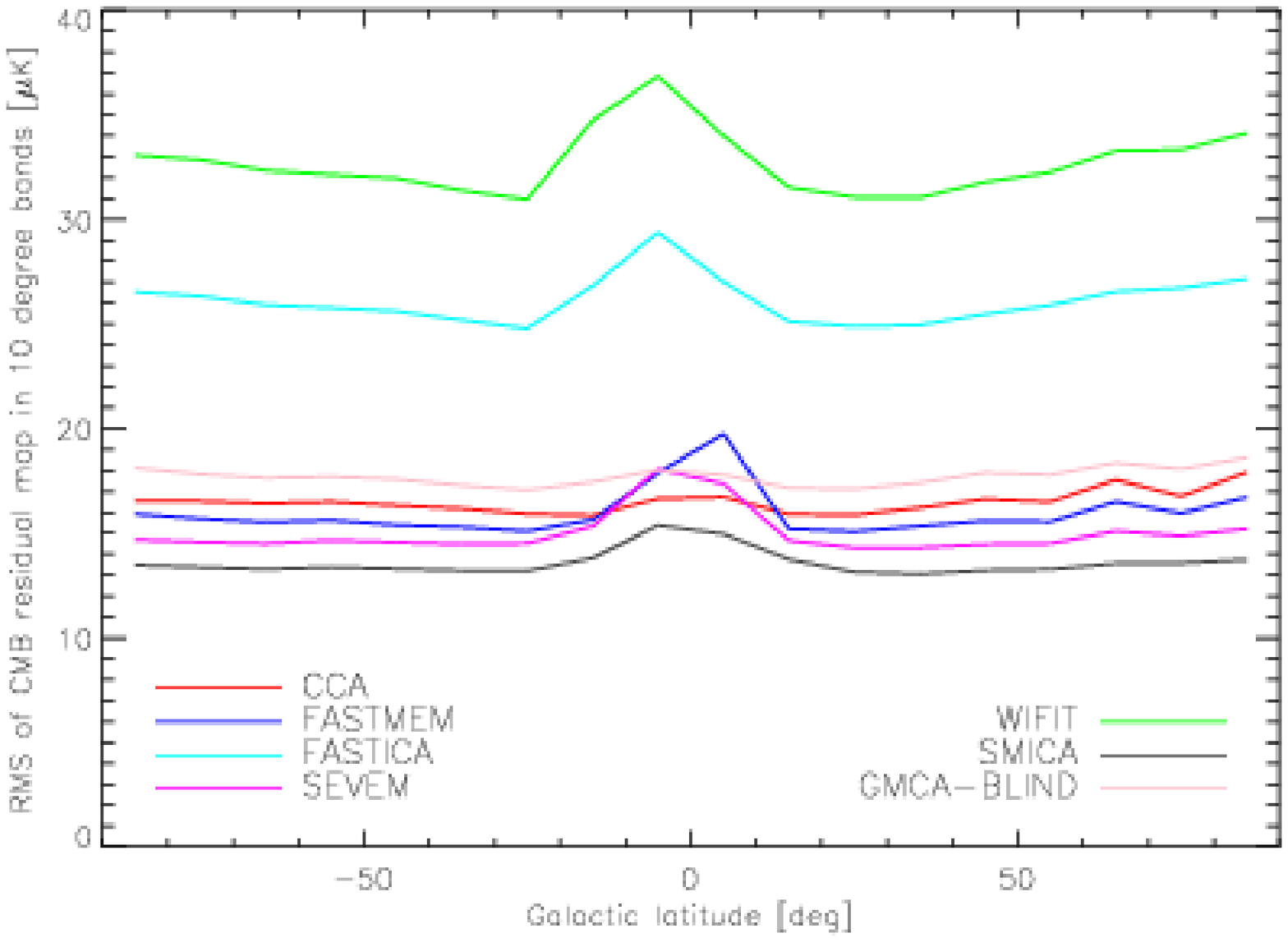}}
  \resizebox{\hsize}{!}{\includegraphics[draft=\draftfig]{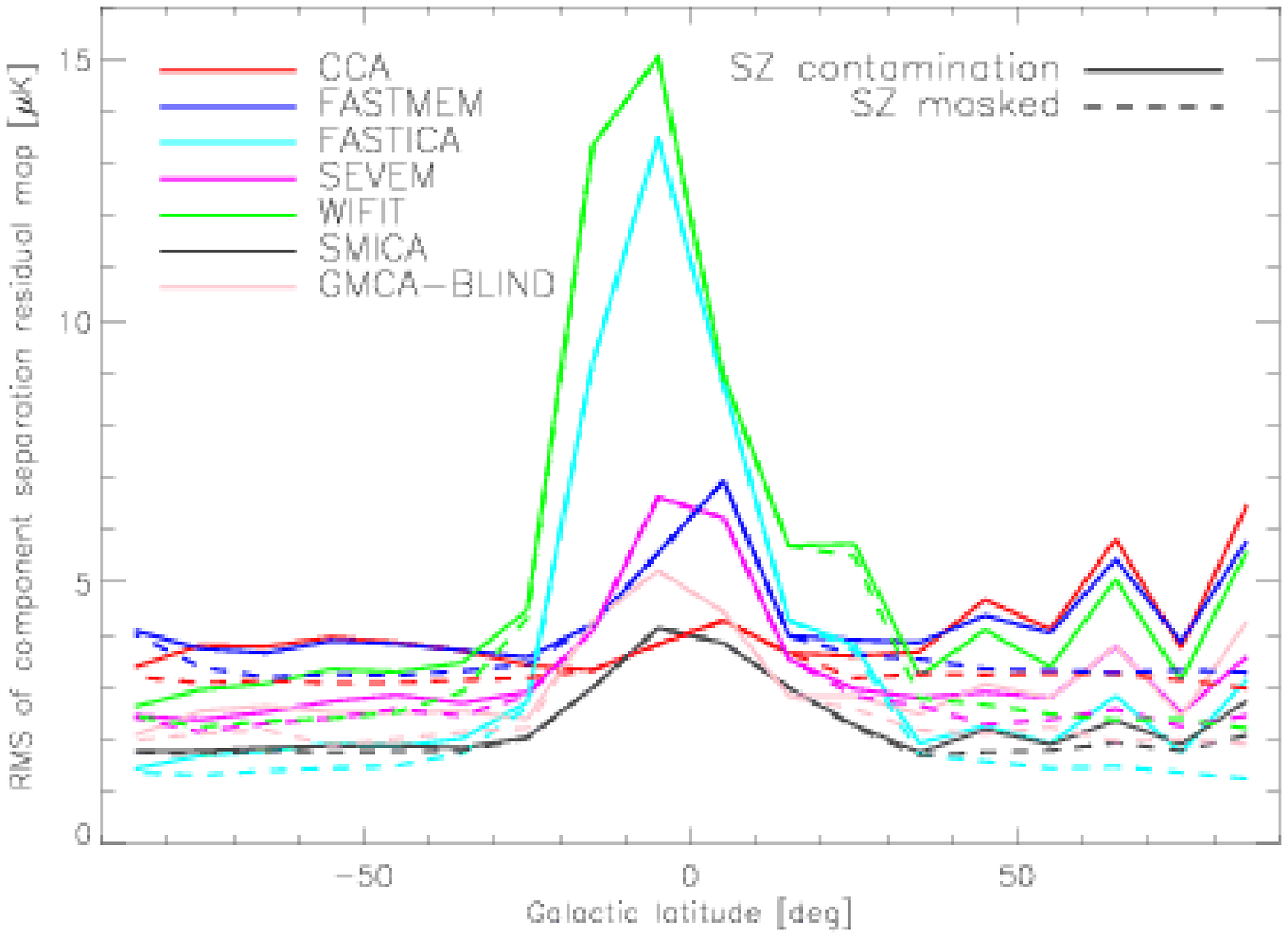}}
  \caption{(Upper) RMS of the residual error of the CMB map,
    calculated for each of 18 bands of 10 degrees width in Galactic
    latitude. For comparison, $\sigma_{{\rm CMB}}=104.5\mu$K and
    $\sigma_{{\rm noise}}=29.3\mu$K, for the 143 GHz channel alone ($1.7'$
    pixels).  (Lower) RMS of this residual map calculated at $45'$
    resolution.  For comparison, $\sigma_{{\rm CMB}}(45')=69.8\mu$K
    and $\sigma_{{\rm noise}}(45')=0.7\mu$K
    for the 143 GHz channel. The corresponding residual maps are shown in
    Figure~\ref{fig:cmbresidualsmoothed_appendix}.}
  \label{fig:cmbresidualspatial}
\end{figure}

\begin{figure*}[]
\centering
\begin{center}
  \begin{tabular}{cc}
    \includegraphics[angle=90,scale=0.35, draft=\draftfig]{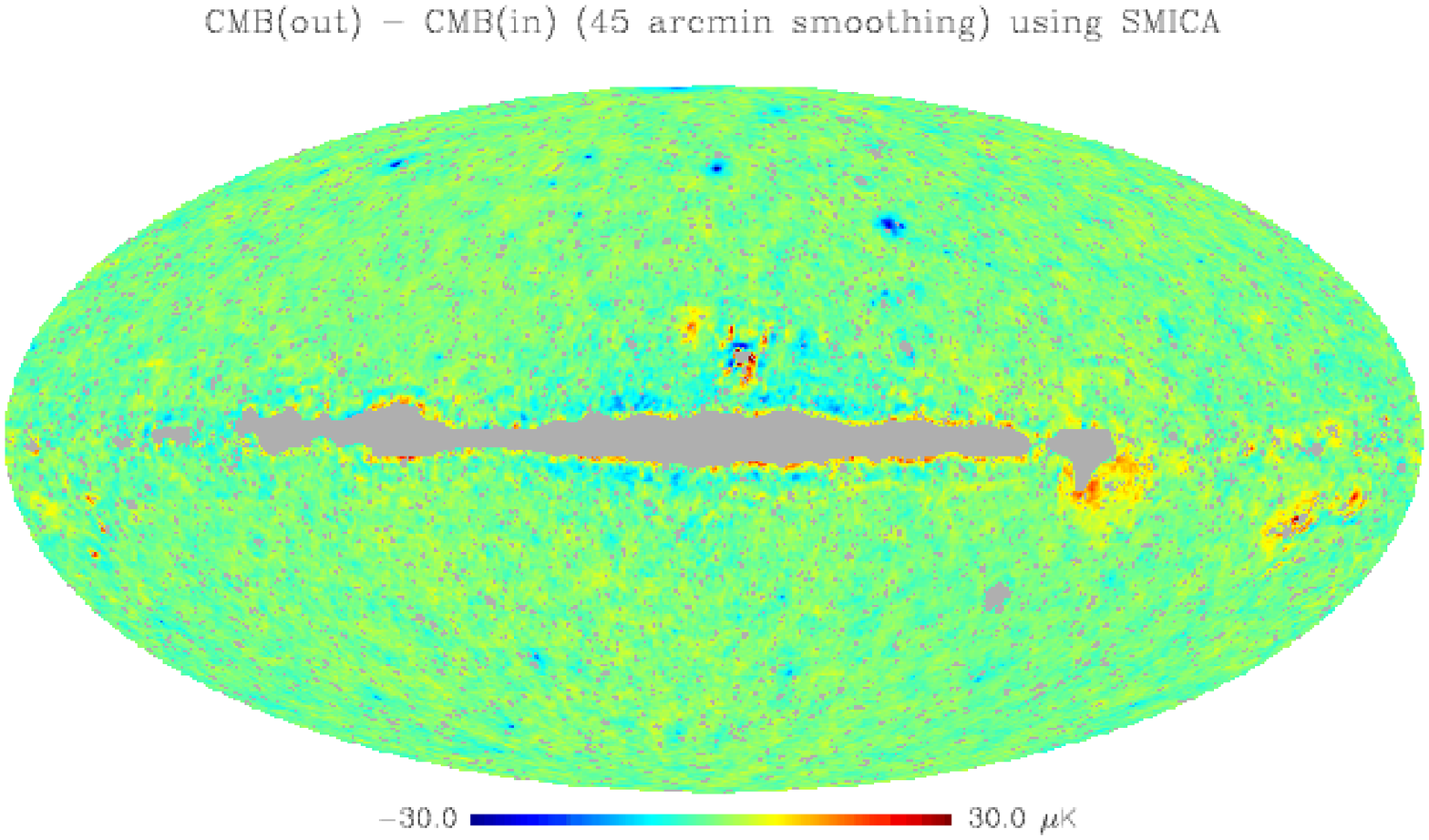}&\includegraphics[angle=90,scale=0.35, draft=\draftfig]{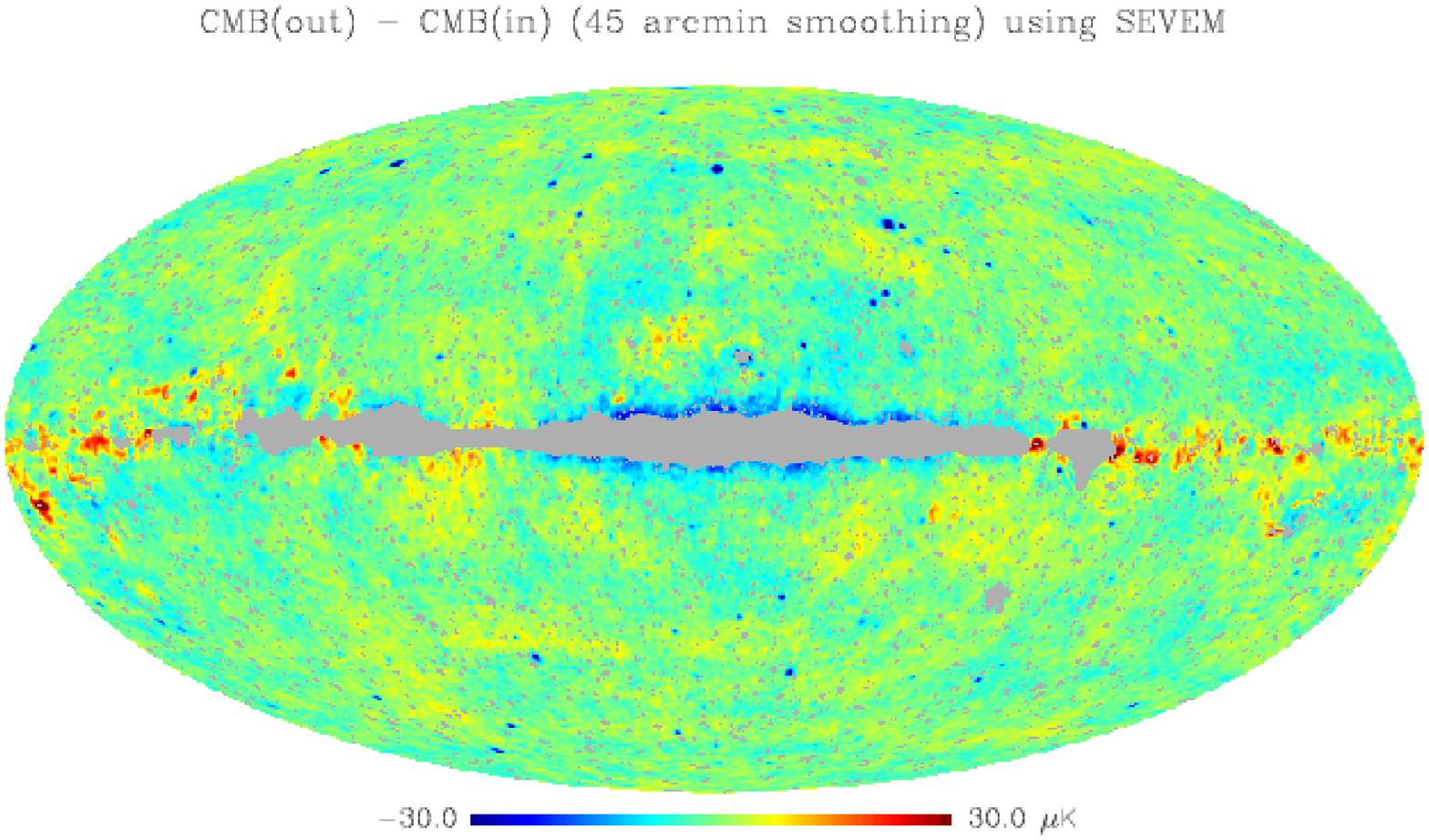}\\
    \includegraphics[angle=90,scale=0.35, draft=\draftfig]{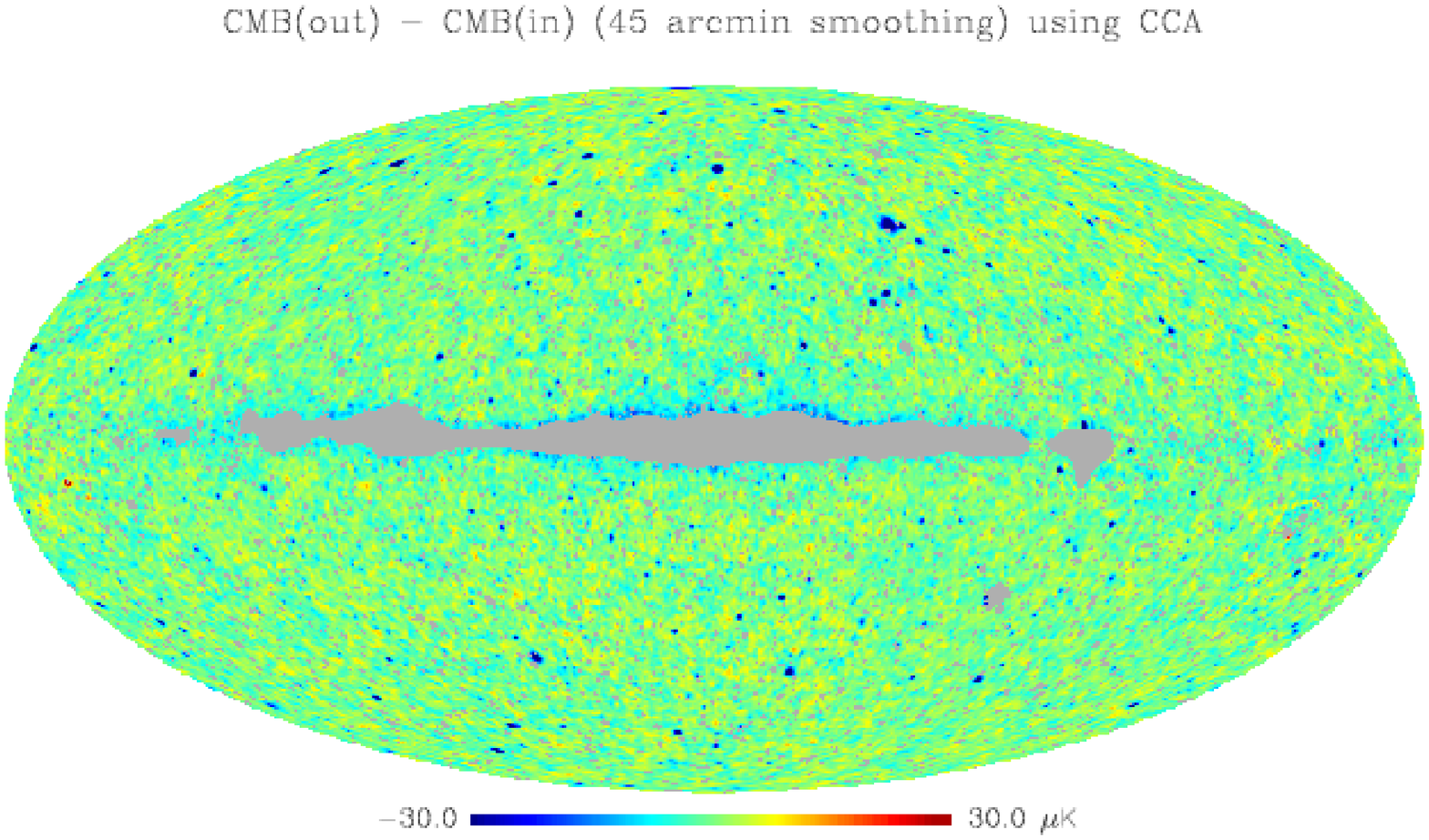}& \includegraphics[angle=90,scale=0.35, draft=\draftfig]{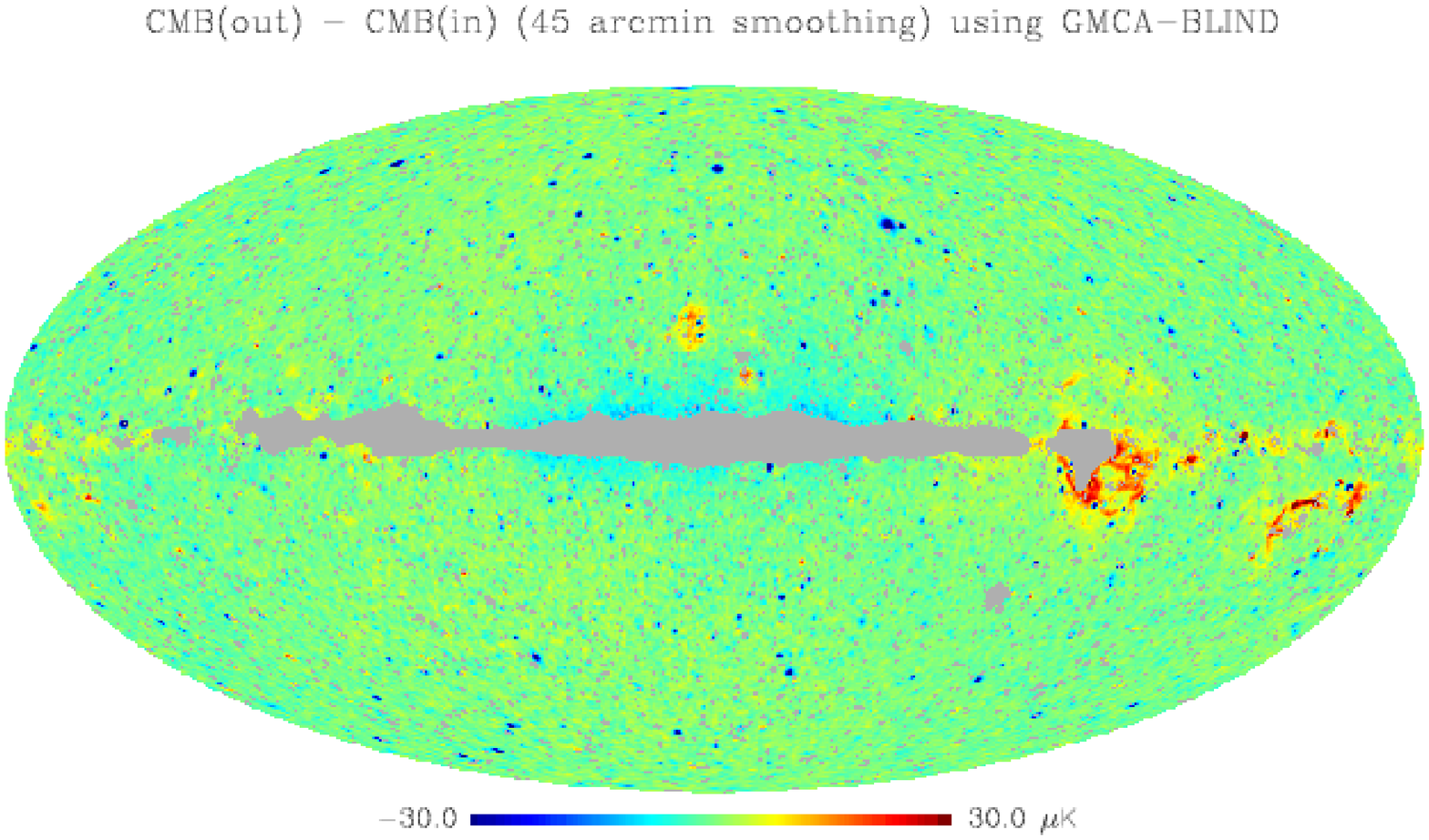}\\
    \includegraphics[angle=90,scale=0.35, draft=\draftfig]{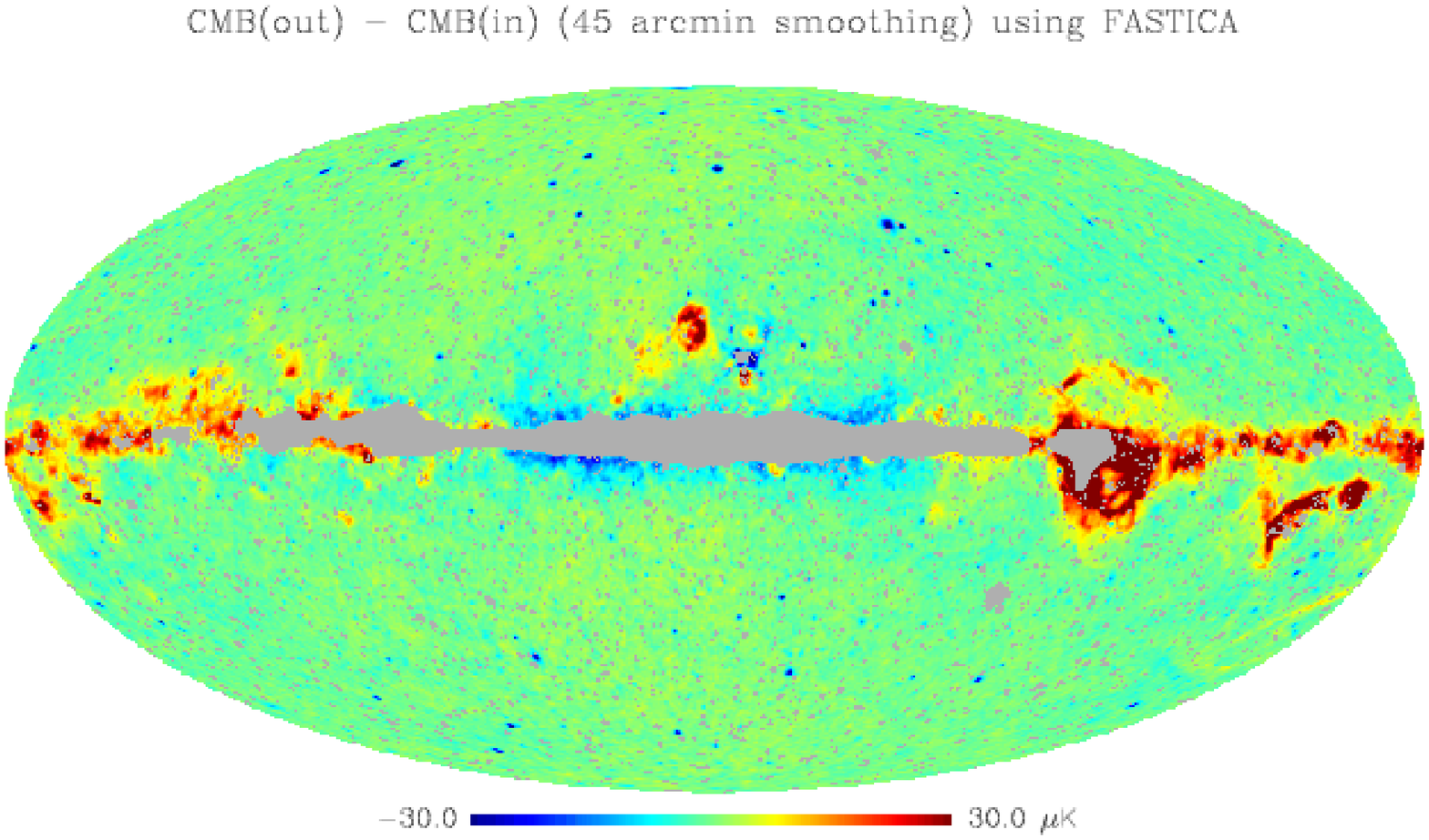}&\includegraphics[angle=90,scale=0.35, draft=\draftfig]{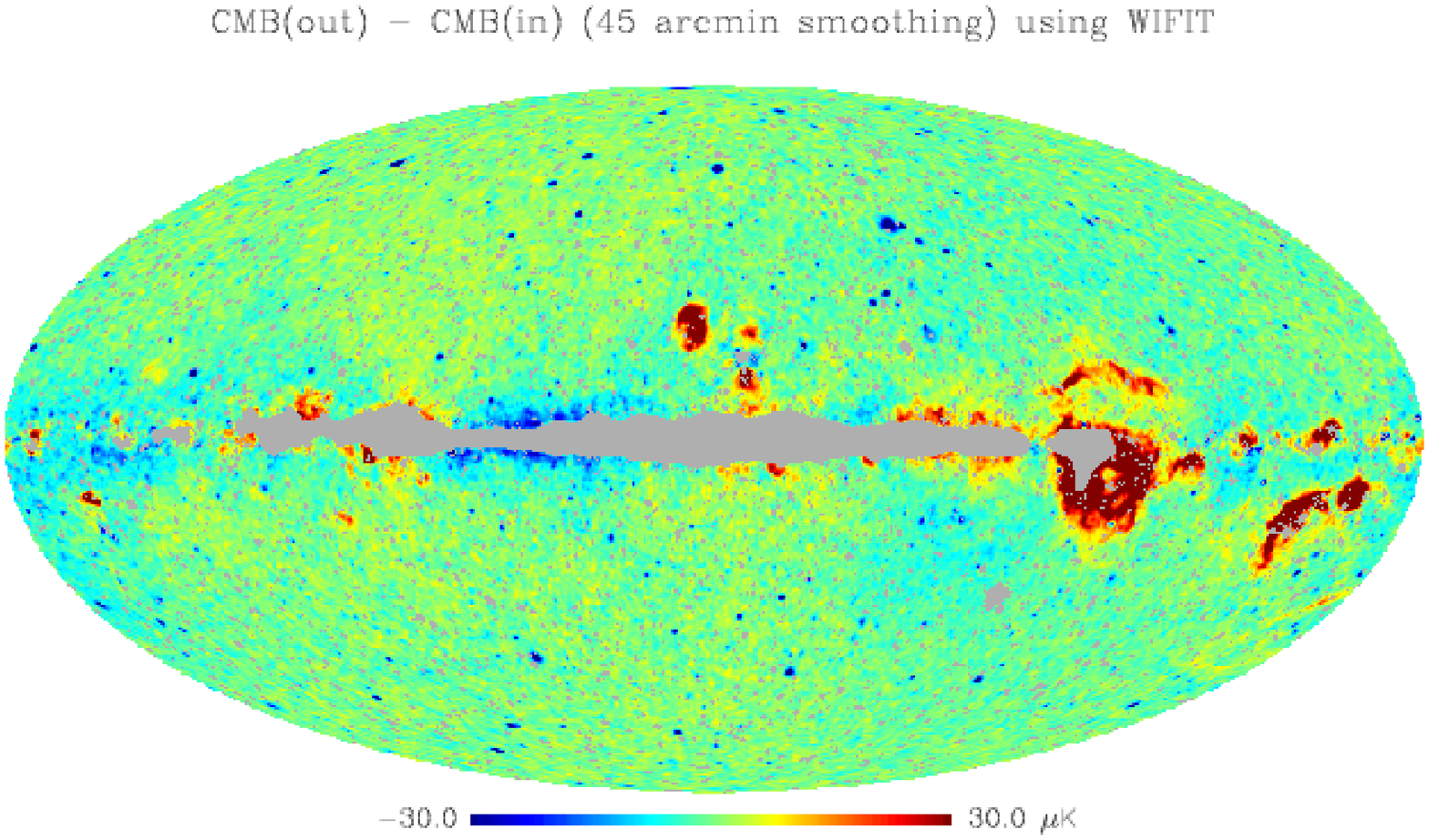} \\
    \includegraphics[angle=90,scale=0.35, draft=\draftfig]{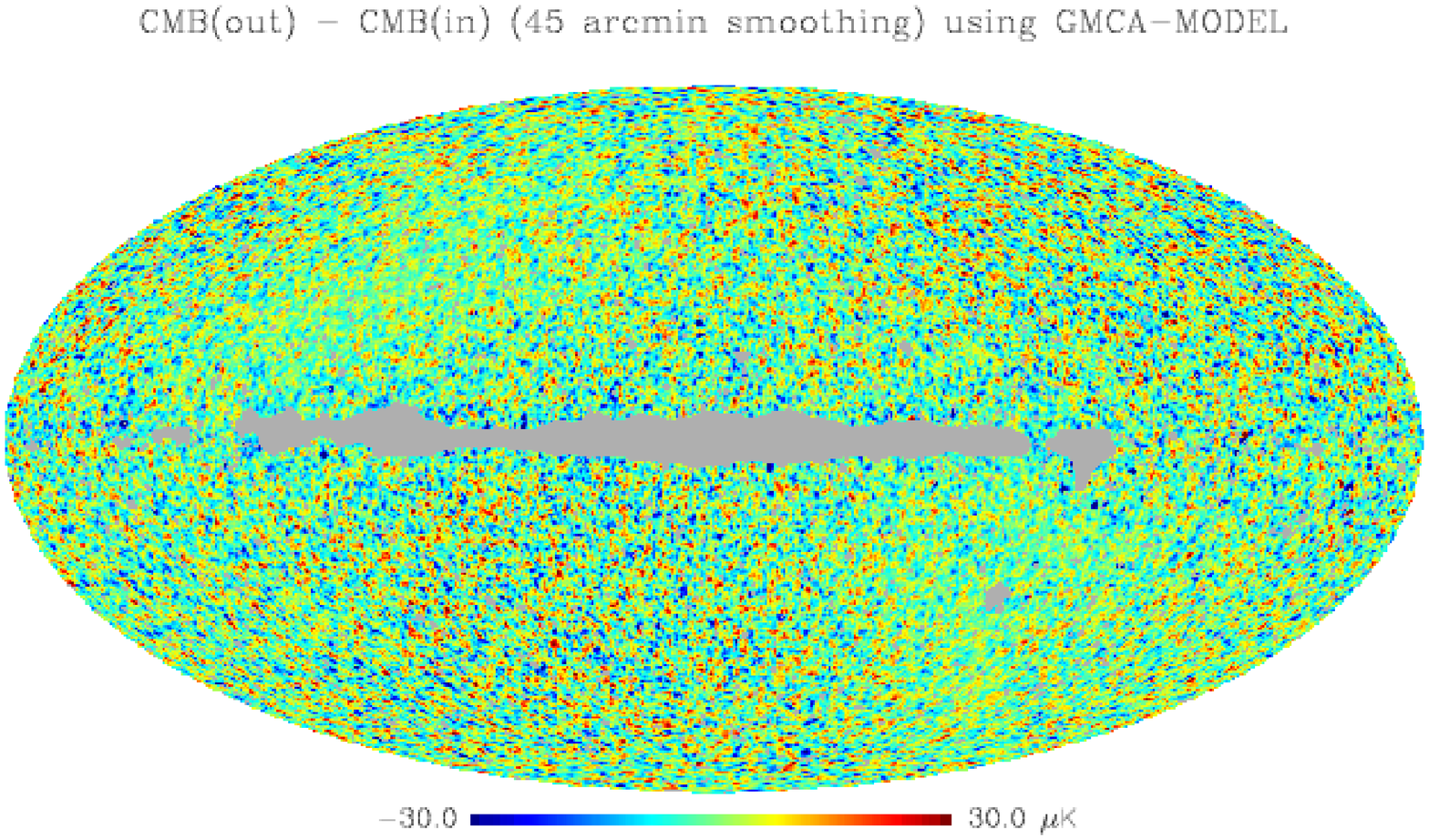} &\includegraphics[angle=90,scale=0.35, draft=\draftfig]{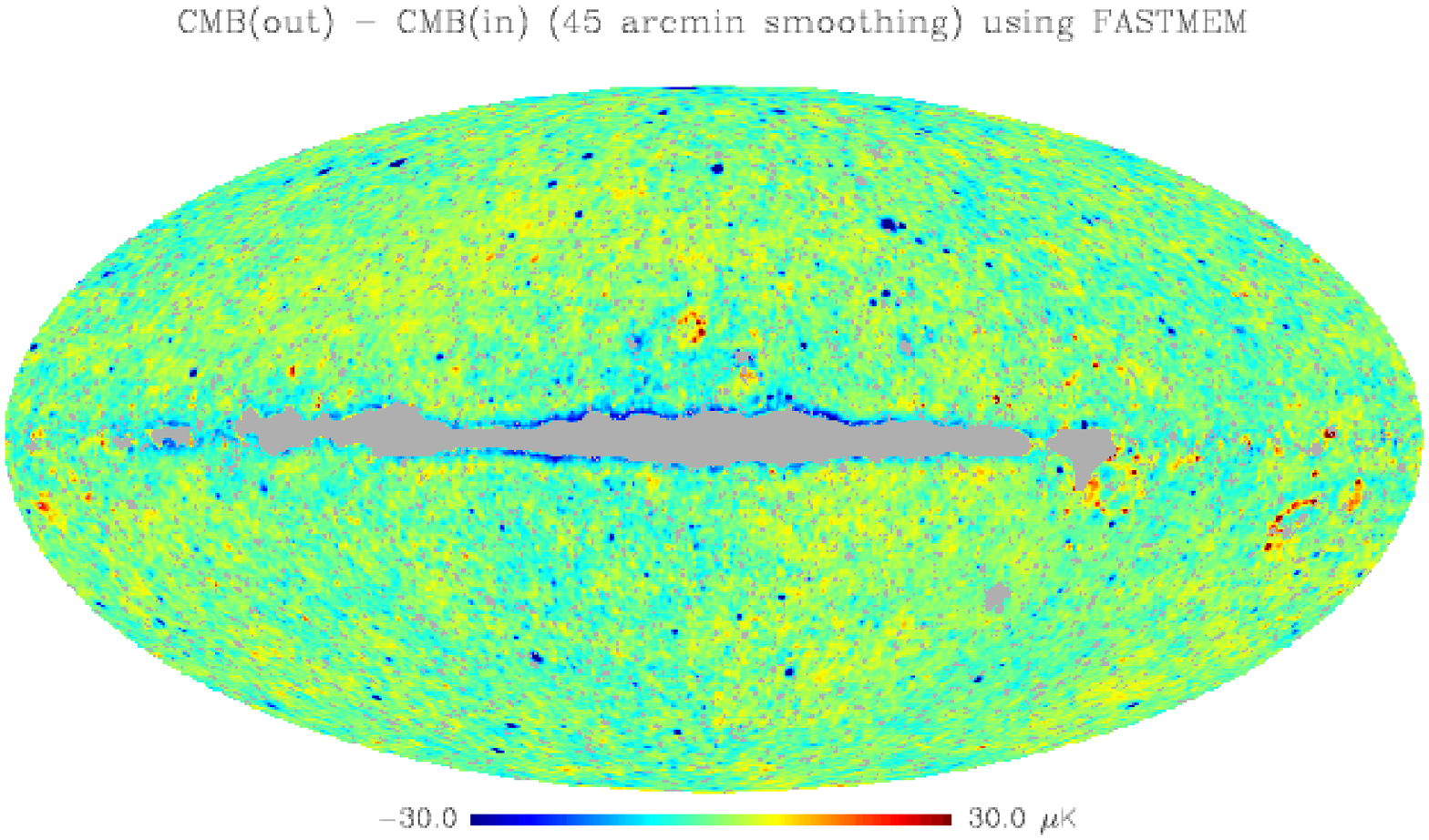}
  \end{tabular}
  \caption[]{{\bf CMB reconstruction error smoothed at $45'$
    resolution}. These maps are described in
    Section~\ref{sec:maplevel}, and their RMS in Galactic latitude
    strips of $10^{\circ}$ are shown in
    Figure~\ref{fig:cmbresidualspatial}.  {{\bfref Over a large
    fraction of sky, the typical RMS of the residual error is between
    2 and 5$\mu$K, which can be compared with $\sigma_{{\rm
    CMB}}(45')=69.8\mu$K and $\sigma_{{\rm noise}}(45')=0.7\mu$K for
    the 143 GHz channel. Some contamination from the galactic components and
    bright clusters remains.}}
  }
  \label{fig:cmbresidualsmoothed_appendix}
\end{center}
\end{figure*}

Maps of the CMB reconstruction error, with all maps smoothed to a common $45'$
resolution, are shown in Figure~\ref{fig:cmbresidualsmoothed_appendix}
for all of the methods (excluding Commander, which produced maps at
3$^{\circ}$ resolution).
The remaining Galactic contamination is now visible at various levels for most
methods, and in particular close to regions with the strongest levels of
free-free emission.  There is also evidence of contamination by SZ
cluster decrements, which are visible as distinct negative sources
away from the Galactic plane.
As can be seen, significant differences between methods exist.
\begin{itemize}
\item At high Galactic latitudes, at this $45'$ scale, the lowest
  contamination is achieved by SMICA, GMCA-BLIND and FastICA.
\item In Zone~2, CCA, GMCA-MODEL, and FastMEM seem to filter out
  Galactic emission best while FastICA and WI-FIT are strongly
  contaminated.
\end{itemize}

A quantitative measure of the raw residual of the CMB map
(reconstructed CMB minus unsmoothed input CMB) is provided by its RMS,
calculated for 18 zonal bands, each 10 degrees wide in Galactic
latitude, excluding pixels in Zone~3 and from the point source mask.
The results are shown in the upper panel of
Figure~\ref{fig:cmbresidualspatial}.  This quantity, denoted
$\sigma_{\Delta {\rm CMB}}$, gives a measure of the sum of the errors
due to residual foreground contamination, noise, as well as from
residual CMB  (due to non unit response on small scales, for
instance).
For orientation, we can see that the ensemble of methods span the
range $13\mu$K$ < \sigma_{\Delta {\rm CMB}}< 35\mu$K, which can be
compared with $\sigma_{{\rm CMB}}=104.5\mu$K and $\sigma_{{\rm
    noise}}=29.3\mu$K, for the 143 GHz channel.  

Similarly, the lower panel of Figure~\ref{fig:cmbresidualspatial} shows
the RMS of the smoothed residual errors shown in
Figure~\ref{fig:cmbresidualsmoothed_appendix}.
Depending on the method, the typical level of foreground contamination
(plus noise) has an RMS from 2 to 5$\mu$K on this smoothing scale.
For comparison, $\sigma_{{\rm CMB}}(45')=69.8\mu$K
and $\sigma_{{\rm noise}}(45')=0.7\mu$K 
for the 143 GHz channel.

\subsection{Spectral residual errors}
\label{sub:spectral-errors}

\begin{figure}[t]
  \centering
  \resizebox{\hsize}{!}{\includegraphics[draft=\draftfig]{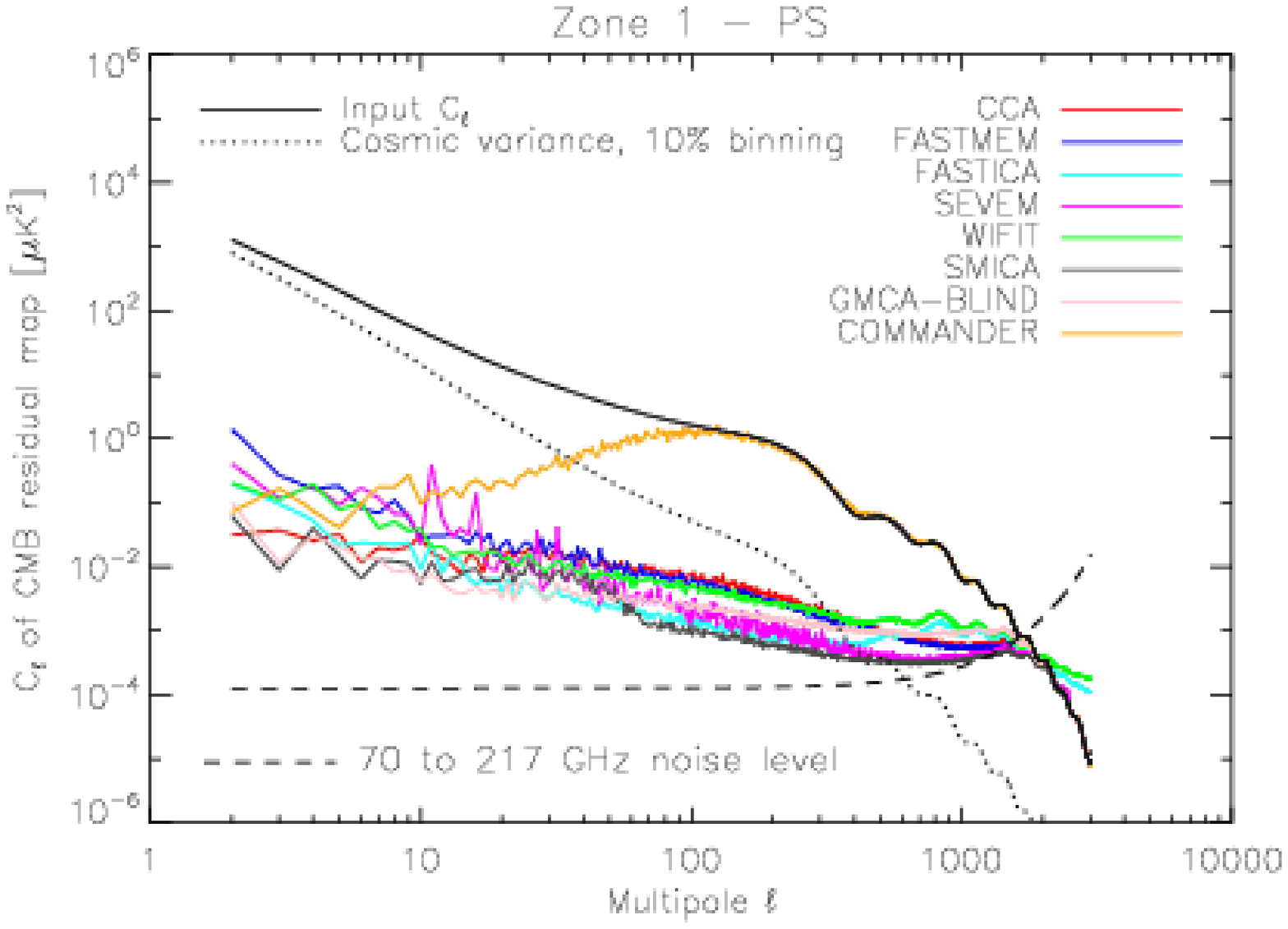}}
  \resizebox{\hsize}{!}{\includegraphics[draft=\draftfig]{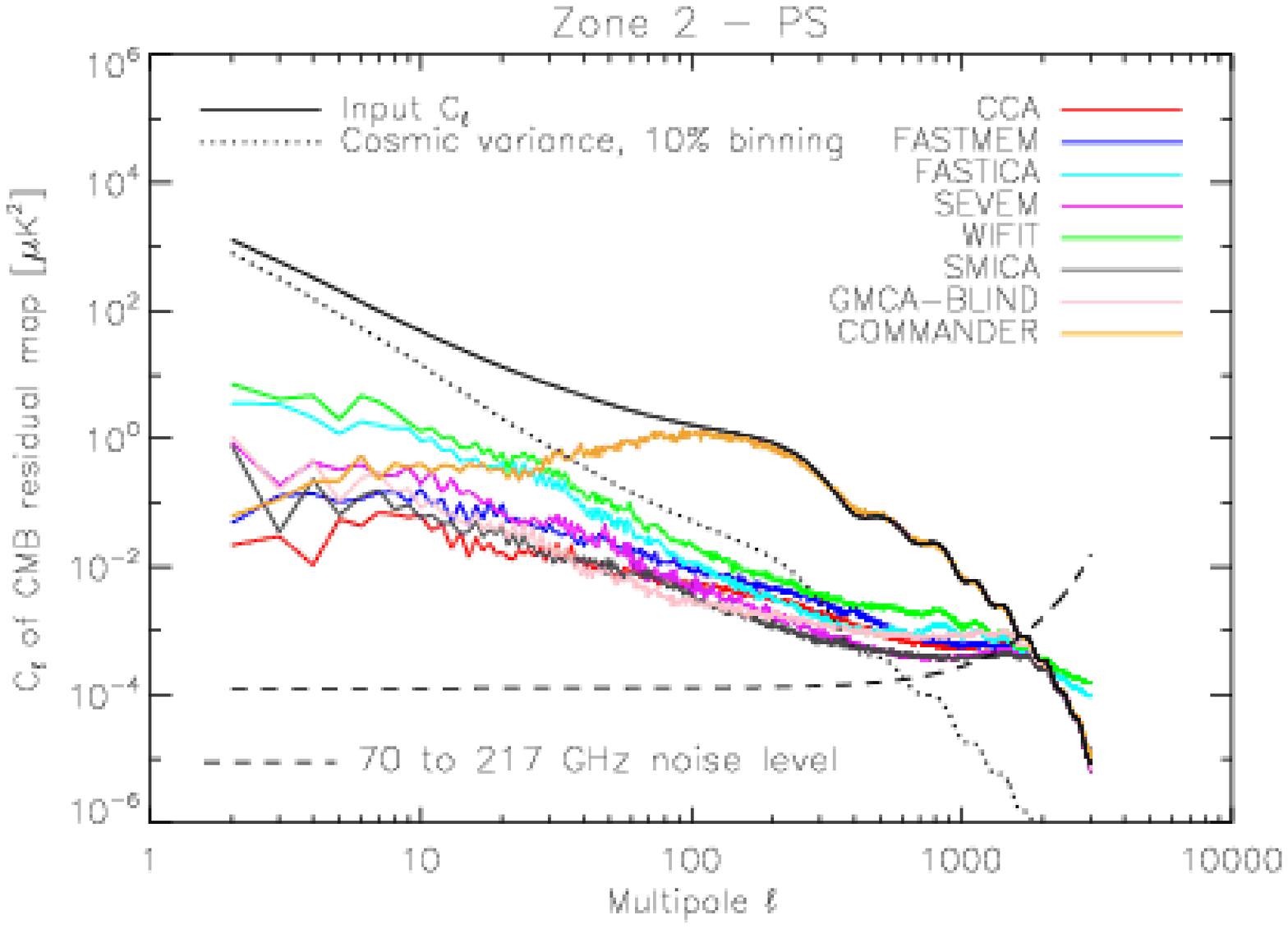}}
  \caption{Spectra of the CMB residual maps, evaluated on Zone~1 (high
    Galactic latitudes) and Zone~2 (low Galactic latitudes), both
    regions with point sources masked. Comparison with
    Figure~\ref{fig:clspectra} shows the extent to which the Galactic
    contamination has been removed from the CMB on large angular
    scales.}
  \label{fig:cmbresidualspectra}
\end{figure}

Next we calculate the spectra of the CMB raw residual maps, both on
Zone~1 and Zone~2 (high and low Galactic latitudes), with the
brightest point sources masked. The results are shown in
Figure~\ref{fig:cmbresidualspectra}.

By comparing the spectra of the residuals with the original level of
diffuse foreground contamination shown in Figure~\ref{fig:clspectra},
we can see that a considerable degree of diffuse foreground cleaning has
been attained.  There seems however to be a `floor' approached by the
ensemble of methods, with a spread of about a factor of ten indicating
differences in performance. This floor appears to be mostly free of
residual CMB signal which would be visible as acoustic oscillations.

Its overall shape is not white: at high Galactic
latitudes the residual spectra bottom out at very roughly $A=0.015
\times\ell^{-0.7} \mu$K$^2$, while a low Galactic latitudes the
spectra bottom out at $A =0.02\times\ell^{-0.9} \mu$K$^2$. This limit
to the level of residuals is considerably higher than the
`foreground-free' noise limit displayed as a dashed line.

It is, however, also significantly lower than the CMB cosmic variance, even
with 10\% binning in $\ell$.  This comforts us in the impression that
component separation is effective enough for CMB power spectrum
estimation (discussed next in this paper), although it may remain a
limiting issue for other type of CMB science. In particular, it
suggests that the component separation residuals, with these channels
and the present methods, will dominate the error in {\sc Planck} CMB
maps.

Recently \citet{Huffenberger:2007sc} performed a reanalysis of the
impact of unresolved point source power in the {\sc WMAP} three-year
data. They found that cosmological parameter constraints are sensitive
to the treatment of the unresolved point source power spectrum beyond
$\ell=200$ characterised by a white noise level of
$A=0.015\pm0.005\mu$K$^2$. By comparison, the residual foreground
contamination obtained in our simulations is as low as
$4\times10^{-4}\mu$K$^2$ at $\ell=200$.

\subsection{Power spectrum estimation errors}

\begin{figure}[]
  \centering
  \resizebox{\hsize}{!}{\includegraphics[draft=\draftfig]{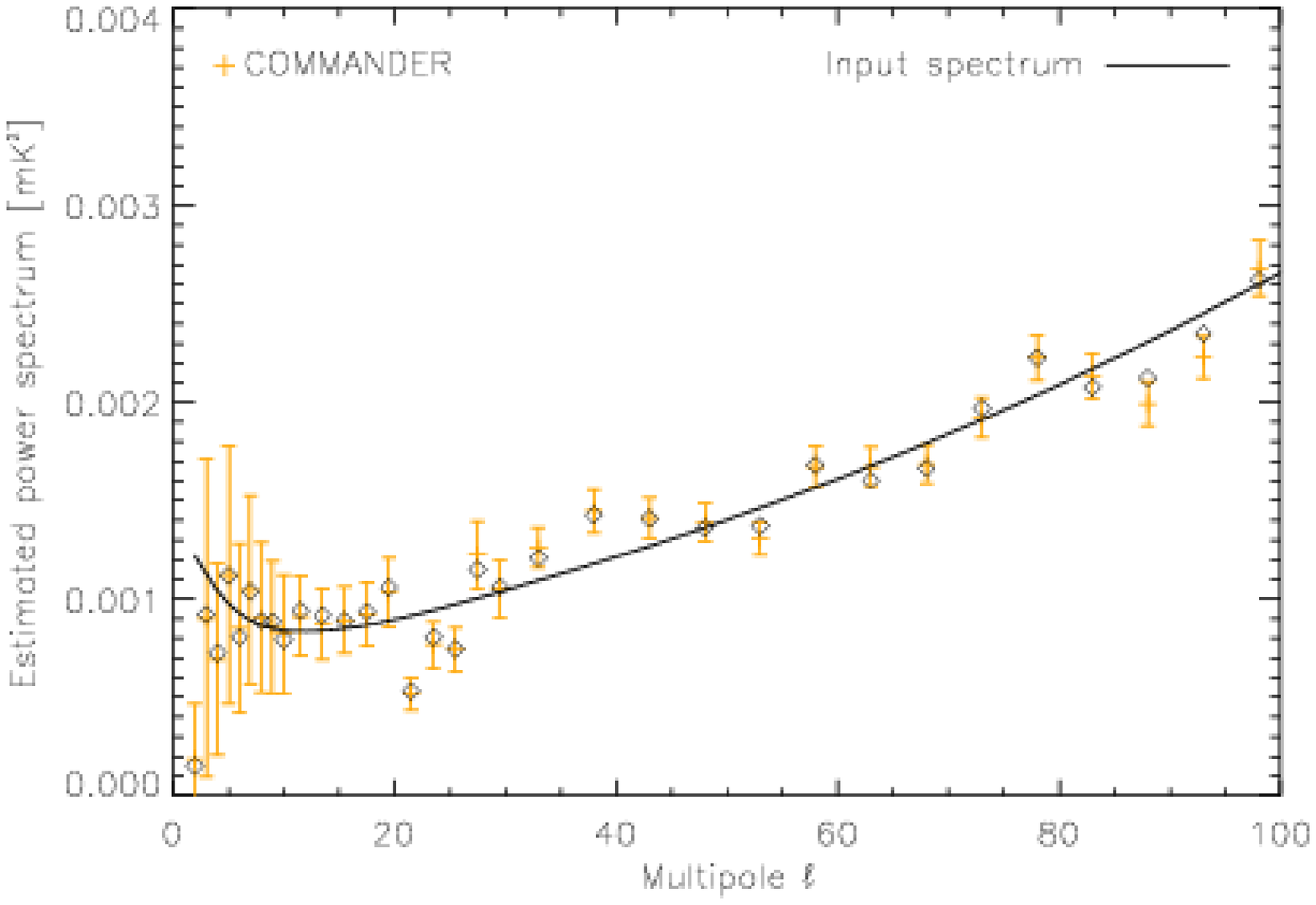}}
  \resizebox{\hsize}{!}{\includegraphics[draft=\draftfig]{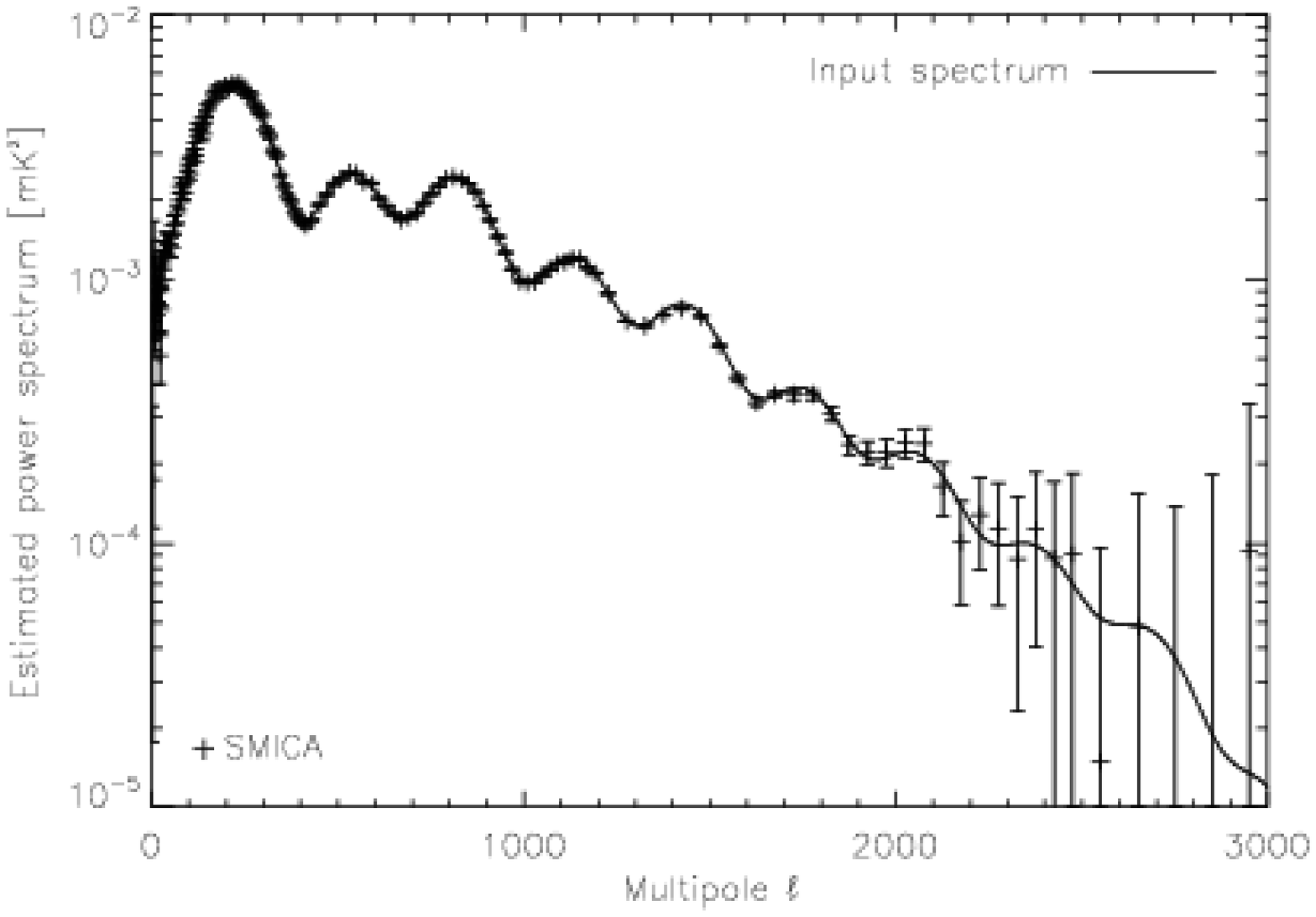}}
  \resizebox{\hsize}{!}{\includegraphics[draft=\draftfig]{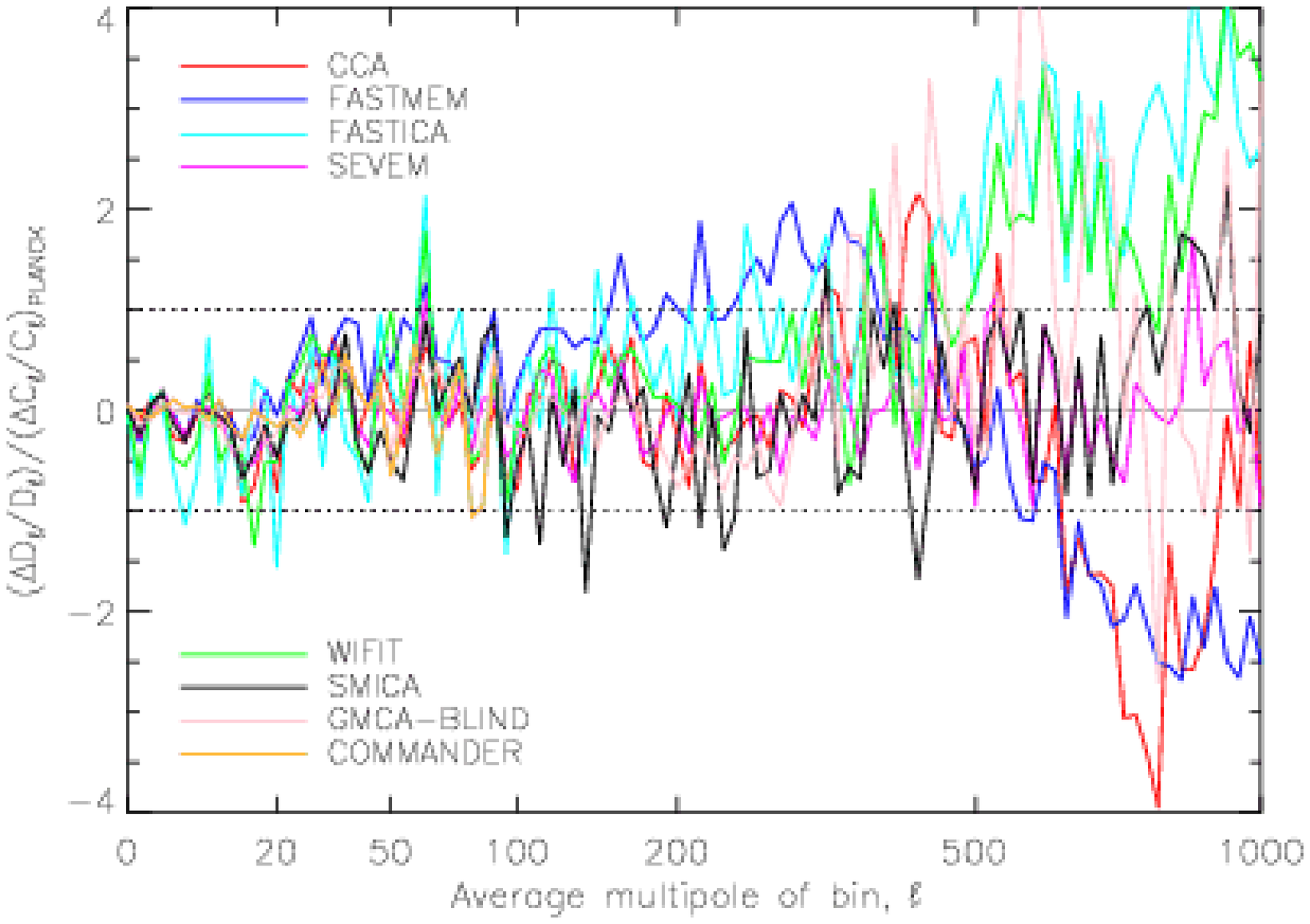}}
  \caption{
    (Upper) Power spectrum estimates (PSE) using Commander on large angular scales.
    The diamonds show the $C_{\ell}$ of the input CMB realisation.
    (Middle) PSE of the recovered CMB map using the SMICA method.  
    (Lower) PSE compared with the estimates derived
    from the input CMB $C_{\ell}$, and with the expected {\sc Planck}
    sensitivity, assuming $f_{\rm sky}=0.8$.  Beyond $\ell=500$ biases
    in the PSE set in in some of the methods.}
  \label{fig:pse_bias}
\end{figure}

Although not the main focus of effort for the Challenge, each group provided
their own bandpower estimates of the CMB power spectrum,
which in many cases showed obvious acoustic structure out to the
sixth or seventh acoustic peak at $\ell\sim2000$.
As an illustration of this result, we show in the upper and
middle panels of Figure~\ref{fig:pse_bias} the power spectrum
estimates from the Commander and SMICA methods respectively.

To make a quantitative estimate of the accuracy of the power spectrum
estimates $D_\ell$ of $\ell(\ell +1)C_{\ell}$, we calculate the
quantity 
\begin{eqnarray}
{\rm FoM}=\frac{\Delta D_{\ell}/D_{\ell}}{\Delta C_\ell/{C_{\ell}}}
\label{psefom}
\end{eqnarray}
where
$\Delta D_{\ell}$ is the bias in the PSE compared to the PSE derived
from the input CMB sky, and where $\Delta C_\ell/C_{\ell}$ is the
expected accuracy of {\sc Planck}, obtained from
Eq.~(\ref{eqn:deltaclovercl}) below.  This figure of merit penalises
biases in the power spectrum estimates without taking into account the
error bars claimed by each group.

In the absence of foregrounds, an approximate lower bound on the relative
standard deviation in estimating the power spectrum is given by
\begin{eqnarray}
  \frac{\Delta C_{\ell}}{C_{\ell}} \simeq \sqrt{\frac{2}{N_{\rm modes}}}\left(1 +\frac{\bar{N}_{\ell}}{C_{\ell}}\right),
\label{eqn:deltaclovercl}
\end{eqnarray}
where the number $N_\mathrm{modes}$ of available modes is
\begin{eqnarray}
  N_{\rm modes} = f_{\rm sky} \sum_{\ell=\ell_{\rm min}}^{\ell_{\rm max}}(2\ell + 1)
\end{eqnarray}
where $f_\mathrm{sky}$ denotes the fraction of sky coverage.
The average noise power spectrum, $ \bar{N}_{\ell}$, is obtained from
the noise power spectra of the different channels
\begin{eqnarray}
  \bar{N}_{\ell} &=& \left(\sum_{\nu =1}^{N_{\rm chan}}\frac{B_{\nu \ell}^2}{N_{\nu \ell}}\right)^{-1},\\
  N_{\nu\ell}&=&\frac{4\pi \sigma_{{\rm hit}}^2}{n_{{\rm pix}}^2}\sum_p\frac{1}{n_{\rm hit}(p)},
\end{eqnarray}
where $B_{\nu\ell}$ is the beam window function for channel $\nu$, and
using the calculated values of $N_{\nu\ell}$ given in Table~\ref{tab:freqchann_Planck}.
This theoretical limit Eq.~(\ref{eqn:deltaclovercl}) is used below to
assess the impact of foregrounds on power spectrum estimation, taking
the 70 to 217GHz channels and assuming the noise levels from
Table~\ref{tab:freqchann_Planck} together with an $f_{\rm sky}=0.8$.

Ideally, the figure of merit given by Eq.~(\ref{psefom}) should be
much less than one in the cosmic-variance limited regime (i.e. for
$\ell\leq500$ according to Figure~\ref{fig:cmbresidualspectra}).
Significant deviations from zero at low $\ell$ and over $\pm1$ at high
$\ell$ are indications of significant departures from optimality.  We
display the FoM Eq.~(\ref{psefom}), calculated for the PSE of each
method in the range $2<\ell<1000$, in the lower panel of
Figure~\ref{fig:pse_bias}.

Focusing first on the range $\ell<20$ we can discern the best
performance from Commander, which models the spatial variation of the
foreground spectral indices, thus improving the subtraction of
foregrounds on large scales. In the range $20<\ell<500$, SEVEM, 
specifically designed for an estimate of the CMB power spectrum,
performs best among the methods tested on this challenge. Beyond
$\ell=500$ we see the best performance from SEVEM and SMICA.

{\bfref At best, it seems that obtaining PSE with ${\rm FoM}<1$ is
achievable for the multipole range $2<\ell<1000$.  Overall though,
complete convergence between the results from different methods was
not yet achieved.}

\subsection{Discussion}

{{\bfref In closing this section on the results for the CMB component,
we make some general observations, and attempt to explain some of the
differences in these results, as shown in
Figure~\ref{fig:cmbresidualsmoothed_appendix}, and with reference to
the characteristics of the analyses as detailed in Table~\ref{tab:methods}.

The most significant foreground contamination, where it exists, is
associated with regions where free-free emission is most intense, such
as the Gum nebula and Orion A and B, and this is most easily visible
in the residual maps of FastICA and WIFIT (which also suffers from
some dust contamination). Possibly this can be explained by the more
limited frequency range of data exploited in these two analyses.

The WMAP data were used in the separation only by Commander and SMICA.
In the Commander case, the inclusion of additional low frequency
channels (in particular the 23 GHz band) helps to recover the low
frequency foregrounds. For SMICA, the use of the WMAP channels was not
really mandatory for extracting the CMB, rather they have been used
with the objective of developing a pipeline which uses all
observations available. WMAP maps are expected to be useful for the
extraction of low frequency galactic components. This specific aspect
was not investigated further in the present work, as the simulations
used here (which do not include any anomalous dust emission) are not
really adequate to address this problem meaningfully.
}}

\section{Results for other components}\label{sec:galcomp}
\subsection{Point sources}

\begin{table*}[t]
  \caption{{\bf Results of point source detection on the present data challenge.}}
  \begin{center}
    \begin{tabular}{c|ccccccccc}
      \hline
      \hline
      Channel    &  30 GHz & 44 GHz & 70 GHz& 100 GHz& 143 GHz& 217 GHz& 353 GHz& 545 GHz& 857 GHz \\
      \hline
      MF: flux limit [mJy] (5\% cont.) & 420 & 430 & 360 & 220 & 130 &100 & 190 & 890 & 2490 \\ 
        Detections & 655 & 591 & 623 & 1103 & 2264 & 2597 & 1994 & 1200 & 1132 \\
      \hline
      MHW2: flux limit [mJy] (5\% cont.) & 395 & 450 & 380 & 250 & 140 & 120 & 230 & 430 & 2160 \\
       Detections & 762 & 621 & 599 & 1065 & 2072 & 2203 & 1650  & 1832 & 1259 \\
      \hline
    \end{tabular}
  \end{center}
  \label{tab:ps}
\end{table*}

Additional efforts have been directed towards producing a catalogue of
{{\bfref point sources}}, a catalogue of SZ clusters and maps of
the thermal SZ effect and Galactic components.

The detection of point sources is both an objective of {\sc Planck}
component separation (for the production of the {\sc Planck} early release
compact source catalogue (ERCSC) and of the final point source
catalogue), and also a necessity for CMB science, to evaluate and
subtract the contamination of CMB maps and power spectra by this
population of astrophysical objects \citep{2008arXiv0803.0577W,2008MNRAS.384..711G}.

The matched filter and the Mexican hat wavelet 2 have advantages and
drawbacks. In principle, the matched filter is the optimal linear
filter.  However, it often suffers from inaccurate estimation of the
required correlation matrix of the contaminants, and from the
difficulty to adapt the filter to the local contamination conditions.
On the data from the present challenge, this resulted in excessive
contamination of the point source catalogue by small scale Galactic
emission, mainly dust at high frequencies.

Table~\ref{tab:ps} summarises the PS detection achieved by these
methods. {{\bfref We found that the Mexican hat wavelet 2 and the
matched filter performed similarly in most of the frequency channels,
and complement each other in the others. Performance depends on the
implementation details, and on properties of the other foregrounds. }}

It should be noted that, for all channels, the 5$\sigma$ detection limit
is somewhat above what would be expected from (unfiltered)
noise alone (by a factor 1.33 for the best case, 44~GHz, to 4.8 for
the worst case, 857~GHz).  This is essentially due to the impact of
other foregrounds and the CMB, as well as confusion with
other sources. In particular, this effect is more evident at 545 and
857~GHz, due to high dust contamination but also due to the confusion with
the highly correlated population of SCUBA
sources~\citep{2004ApJ...600..580G,2004MNRAS.352..493N}, which
constitute a contaminant whose impact on point source detection was
until now somewhat underestimated.

The number of detections for each frequency channel in
Table~\ref{tab:ps} has been compared to the predictions made by
\citet{lcaniego06}, properly rescaled for our sky coverage. In
general, there is a good agreement between the predictions and the
results of this exercise, except for the 857~GHz channel, where the
number of detections is roughly half the predicted one. Again, the
difference may be due to the confusion of correlated infrared sources,
that are now present in the PSM but were not considered by
\citet{lcaniego06}.

\subsection{SZ effect}
\label{sec:szresults}

\begin{figure*}
  \begin{center}
    \begin{tabular}{cc}
      \includegraphics[scale=0.5]{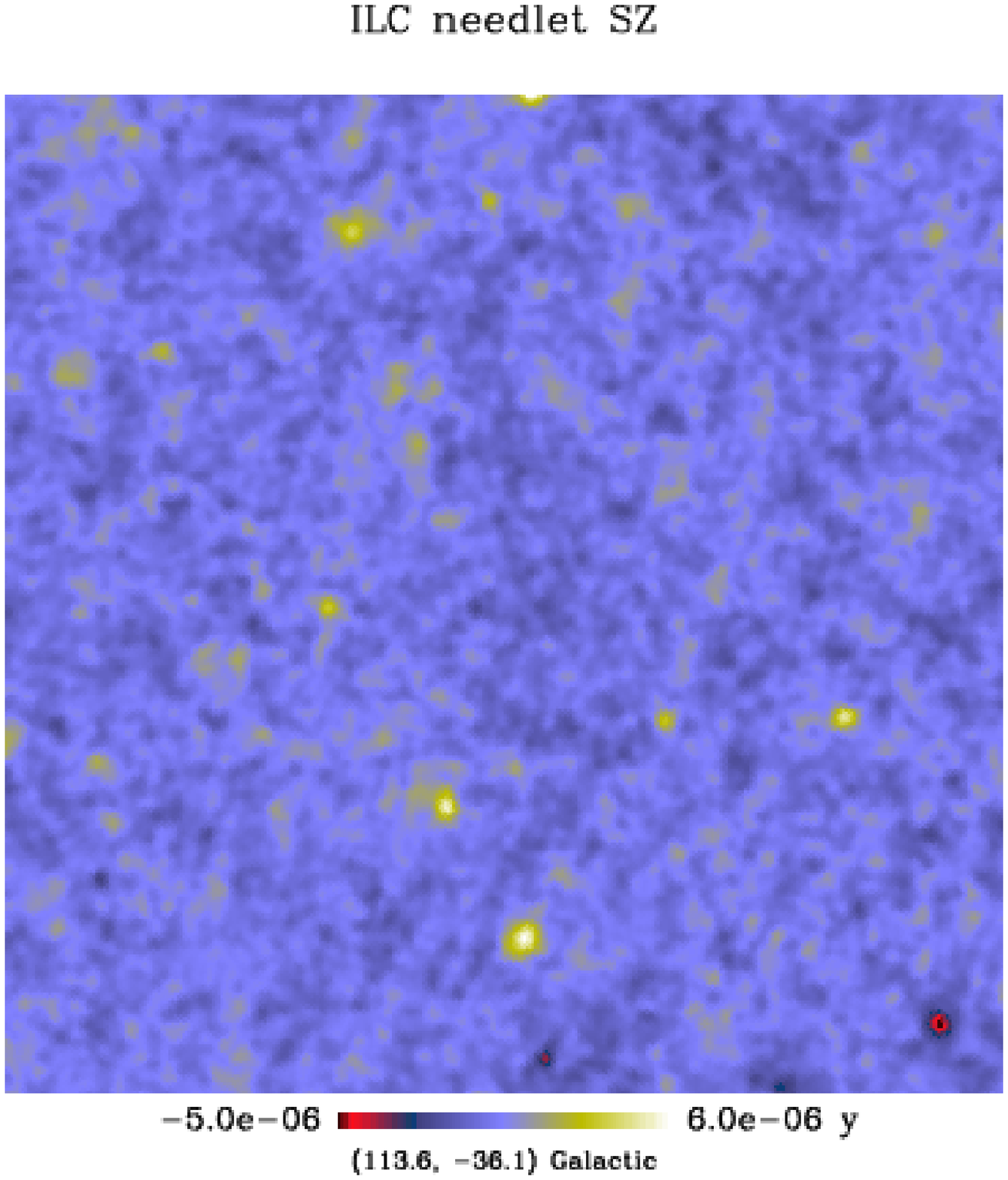} & \includegraphics[scale=0.5]{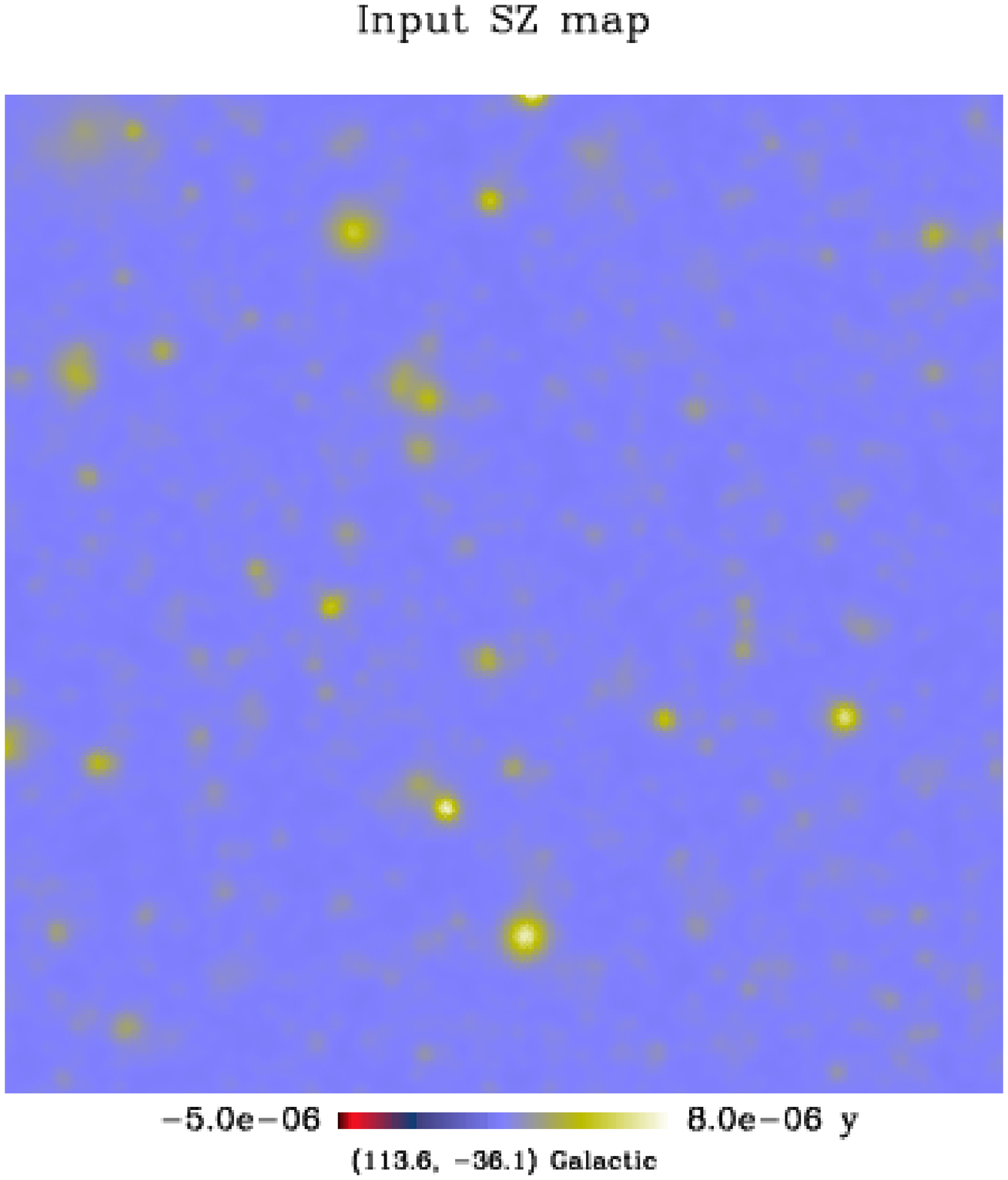}
    \end{tabular}
    \caption[]{{\bf Patch of the recovered needlet-ILC SZ map and input SZ map.} For easier comparison of the two maps (12.5~deg $\times$ 12.5~deg), the input SZ map has been filtered to the same resolution as the output.}
    \label{fig:sz_patches}
  \end{center}
\end{figure*}

The recovery of an SZ map from the challenge data is illustrated in
Figure~\ref{fig:sz_patches}.  The recovered full sky SZ is obtained by
Wiener-filtering in harmonic space the needlet-ILC map of the SZ $y$
parameter.  Wiener filtering enhances the visibility of SZ clusters.
We clearly identify by eye the brightest clusters in the map. 

One of the main results of this study is the recovery of around 2300
clusters.  This is significantly lower than the performance one could
expect if the main limitation was the nominal {\sc Planck} noise, and if
most detectable clusters were unresolved.  Many of the recovered
clusters are in fact resolved, and thus emit on scales where the
contamination from CMB is not negligible.  Small scale Galactic
emission and the background of extragalactic sources, now included in
the simulations, further complicate the detection.  Further study is
necessary to find the exact origin of the lack of performance, and
improve the detection methods accordingly.

Actual detection performance, limited to 67\% of the sky at Galactic
latitudes above 20 degrees, is shown in
Table~\ref{tab:sz_performances}. The ILC~+~SExtractor method gives the best result.
The ILC+MF approach performs as well
as the matched multifilter here. The two implementations of the MMF
perform similarly. The difference in the number of detections achieved
(about 13.5\%), however, suggests that implementation details are
important for this task.

Using the detected cluster catalogue obtained with the MMF, we
have produced a mask of the detected SZ clusters. For each of the 1625
clusters we masked a region whose radius is given by the corresponding input
cluster virial radius (ten times the core radius here).

\begin{table}
      \caption[]{{\bf Performance of the SZ cluster detection
      methods.} The table gives the absolute number of detection for
      $|b|>20^{\circ}$.
      }
         \label{tab:sz_performances}
        \begin{center}
        \begin{tabular}{c|c|c|c}
         \hline
         \hline
          Method    &  Detections & False & Reliability \\
         \hline
         Needlet-ILC + SExt. & 2564 & 225 & 91\% \\ 
         Needlet-ILC + MF & 1804 & 179 & 90\% \\ 
         MMF Saclay & 1803 & 178 & 90\% \\ 
         MMF IFCA & 1535 & 144 &  91\% \\
        \hline
        \end{tabular}
        \end{center}
\end{table}

\subsection{Galactic components}\label{sec:galresults}

For the Challenge, a number of methods were applied for separating out
Galactic components.  Table~\ref{tab:methods} lists which Galactic
components were obtained by the different methods.  Five groups have
attempted to separate a high frequency dust-like component.  Four
groups have attempted separation of synchrotron and free-free at low
frequencies.

We compared the reconstructed component maps with their counterpart
input maps, both in terms of the absolute residual error and in terms
of the relative error.  Both these measures are computed after
removing the best-fit monopole and dipole from the residual error map (fitted
when excluding a region $\pm30^{\circ}$ in Galactic latitude). 
We then defined a figure of merit $f_{20\%}$, which corresponds to the
fraction of sky where the foreground amplitude is reconstructed with a
relative error of less than 20$\%$.

The main results can be summarised as follows:
The dust component was the best reconstructed component with
an $f_{20\%}\simeq0.7$ for all methods. The relative error typically
becomes largest at the higher galactic latitudes where the dust emission
is faintest.
The synchrotron component was reconstructed with an $f_{20\%} \simeq
0.3$--$0.5$, with Commander achieving the best results at $3^{\circ}$
resolution, but with noticeable errors along the galactic ridge where,
in our simulations, the synchrotron spectral index flattens off.
Free-free emission is detected and identified in regions such as the
Gum Nebula, Orion A and B, and the Ophiucus complex.  However, the
reconstruction of the free-free emission at low Galactic latitudes
needs improving.  
On the other hand, the \emph{total} Galactic emission (free-free plus
synchrotron) at low-frequencies is better reconstructed, with
$f_{20\%}\simeq0.5$--$0.8$, with the best results from Commander.

\begin{figure*}
   \centering
    \begin{center}
    \begin{tabular}{cc}
    \includegraphics[angle=90, scale=0.31]{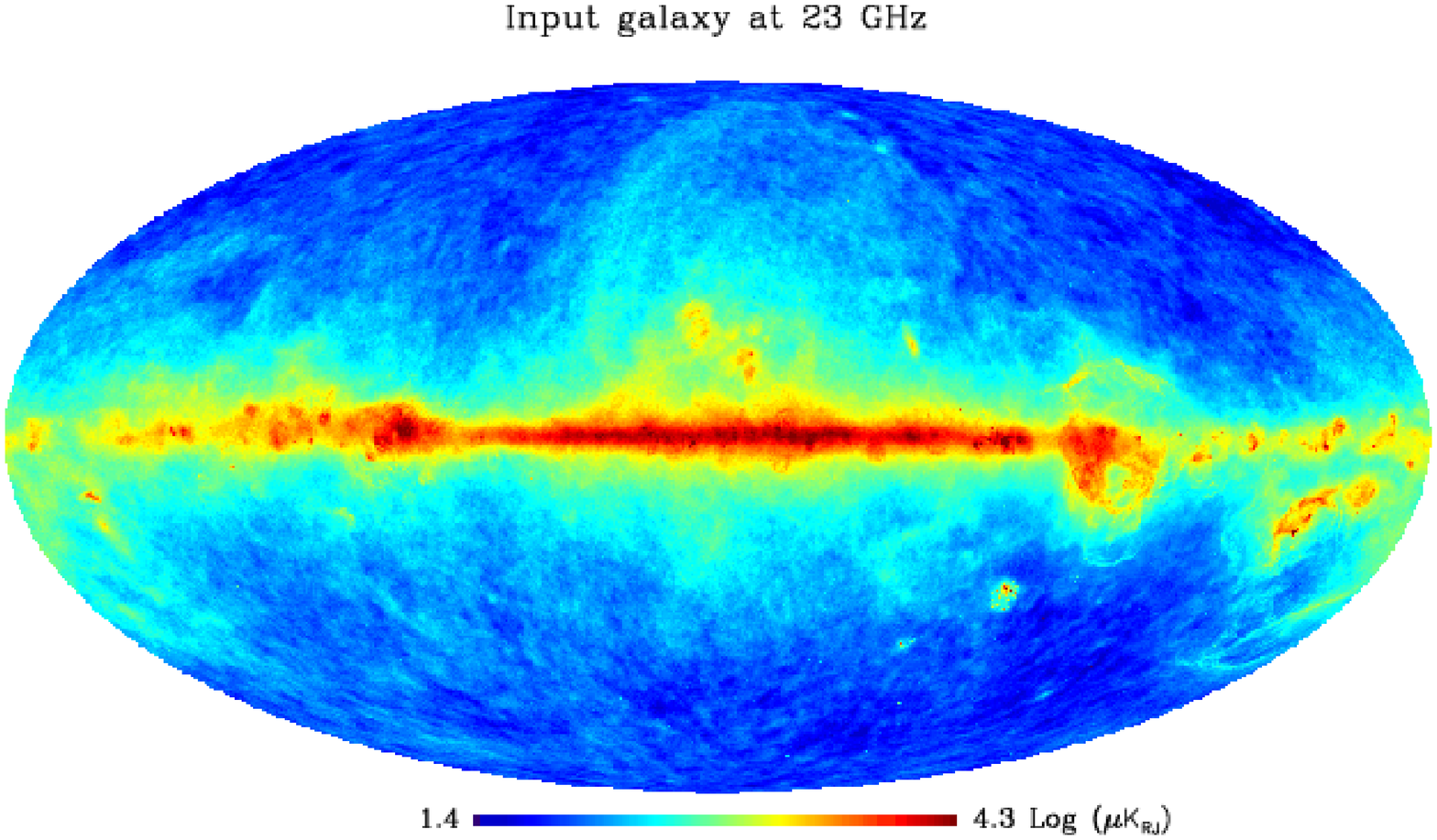} & \includegraphics[angle=90, scale=0.31]{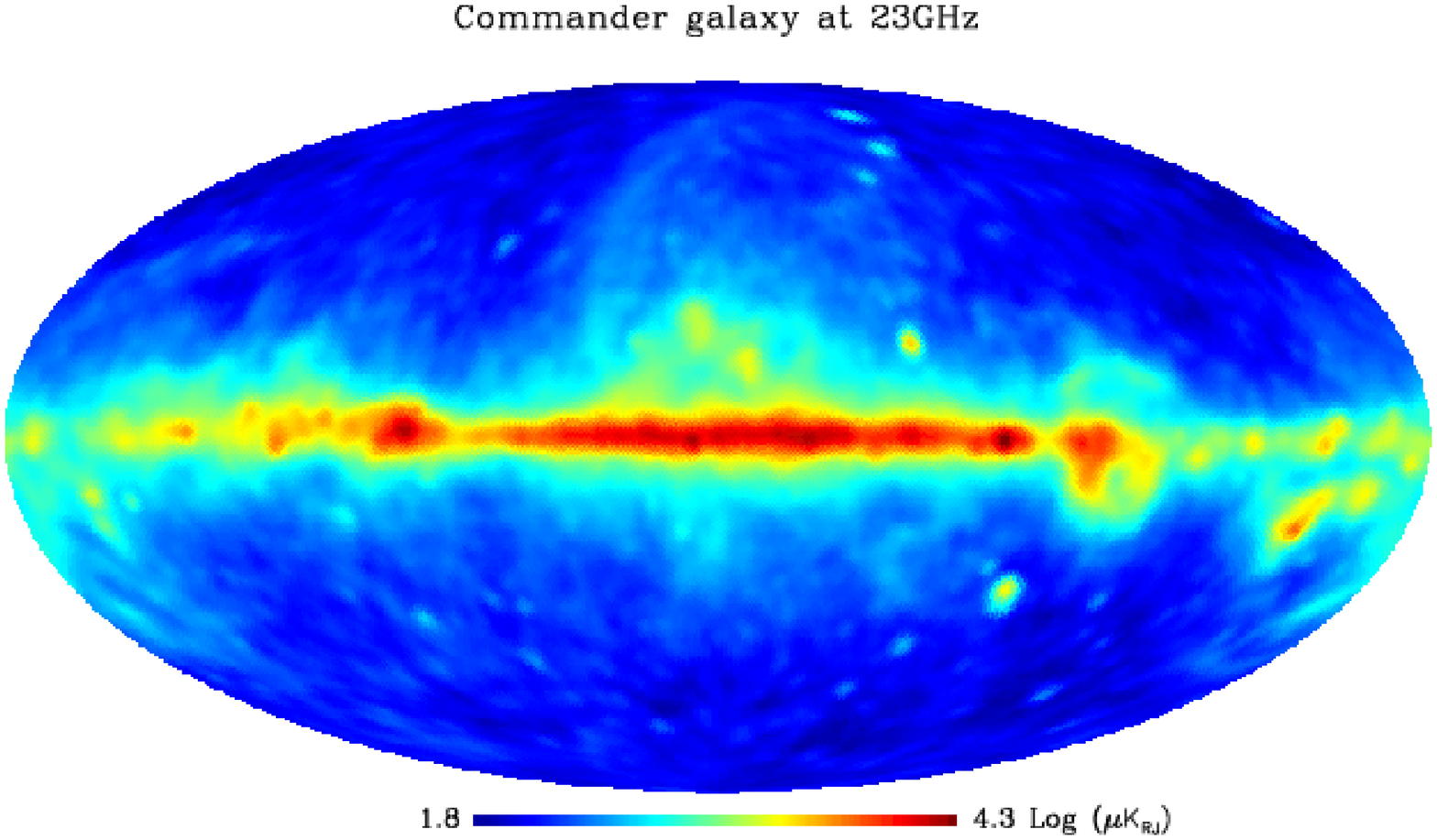}\\
    \includegraphics[scale=0.5]{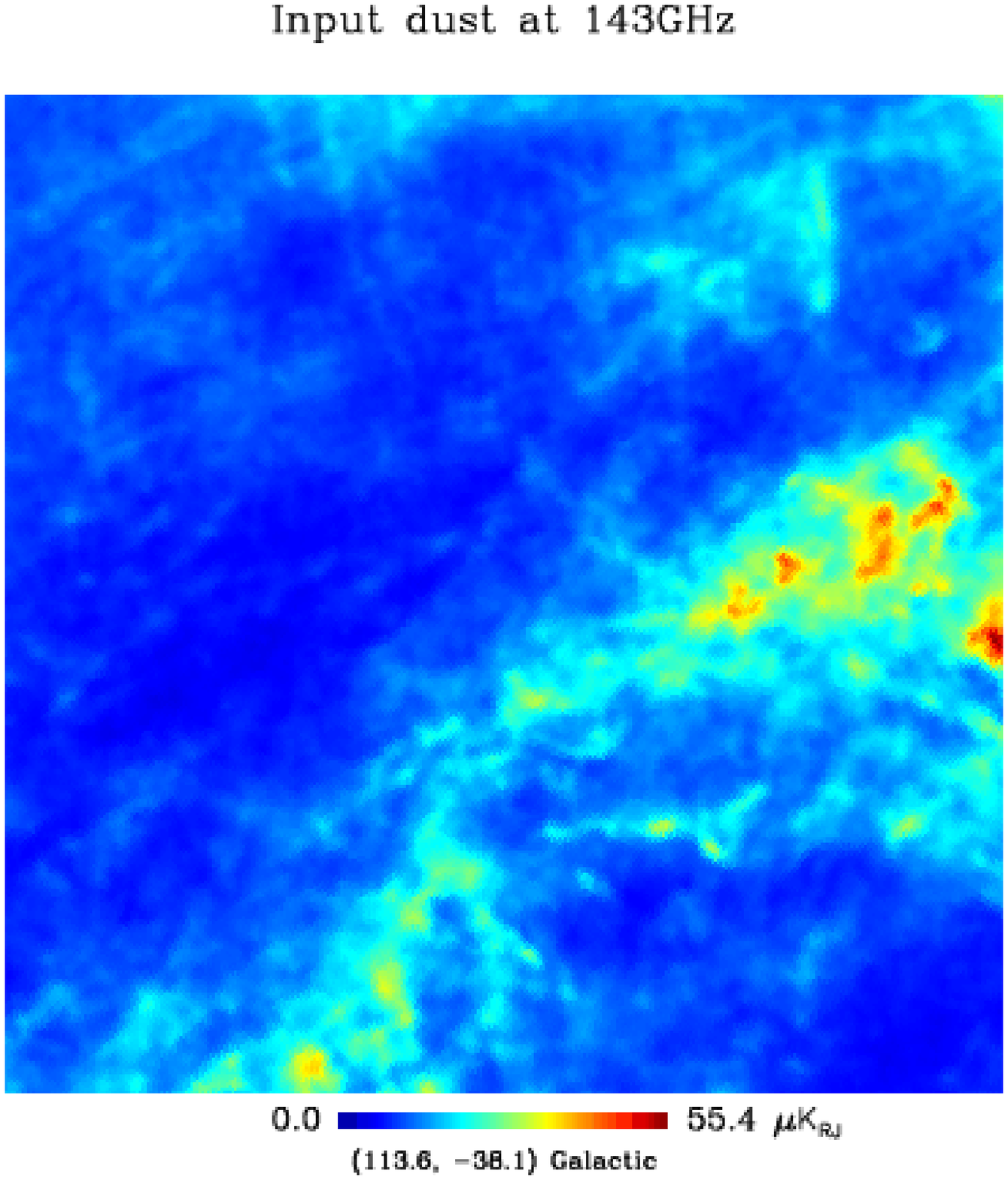} & \includegraphics[scale=0.5]{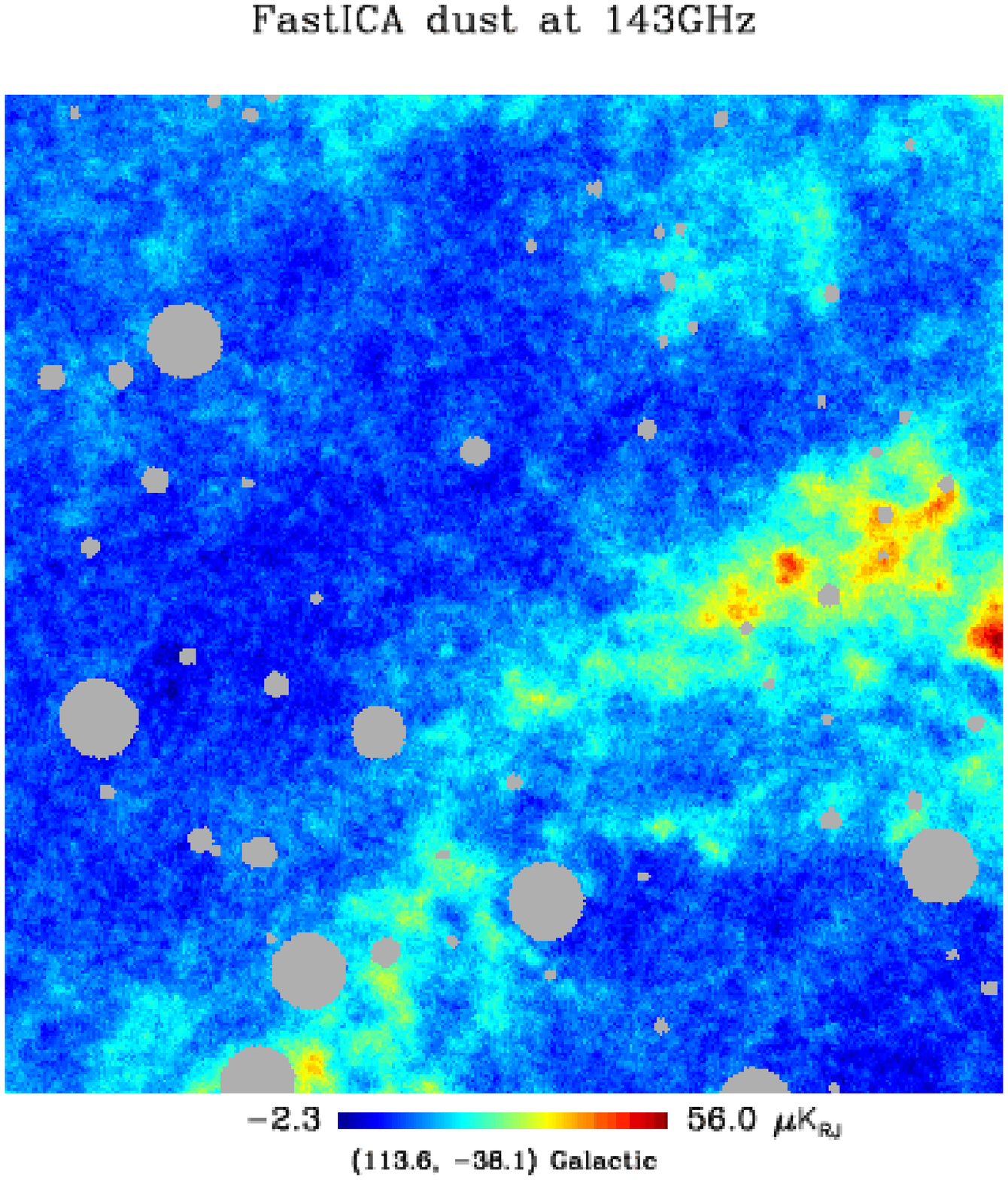}\\
    \includegraphics[scale=0.5]{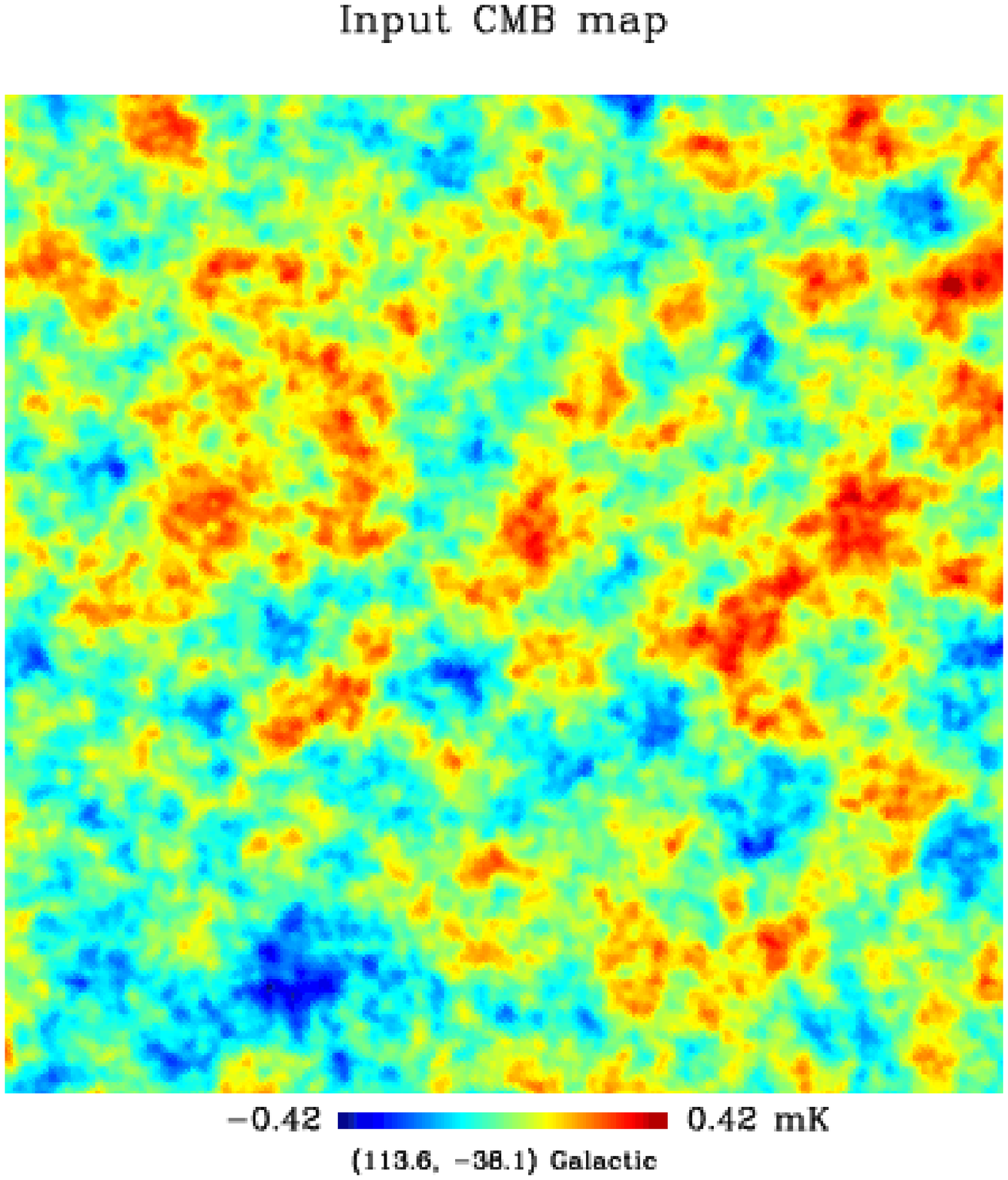} & \includegraphics[scale=0.5]{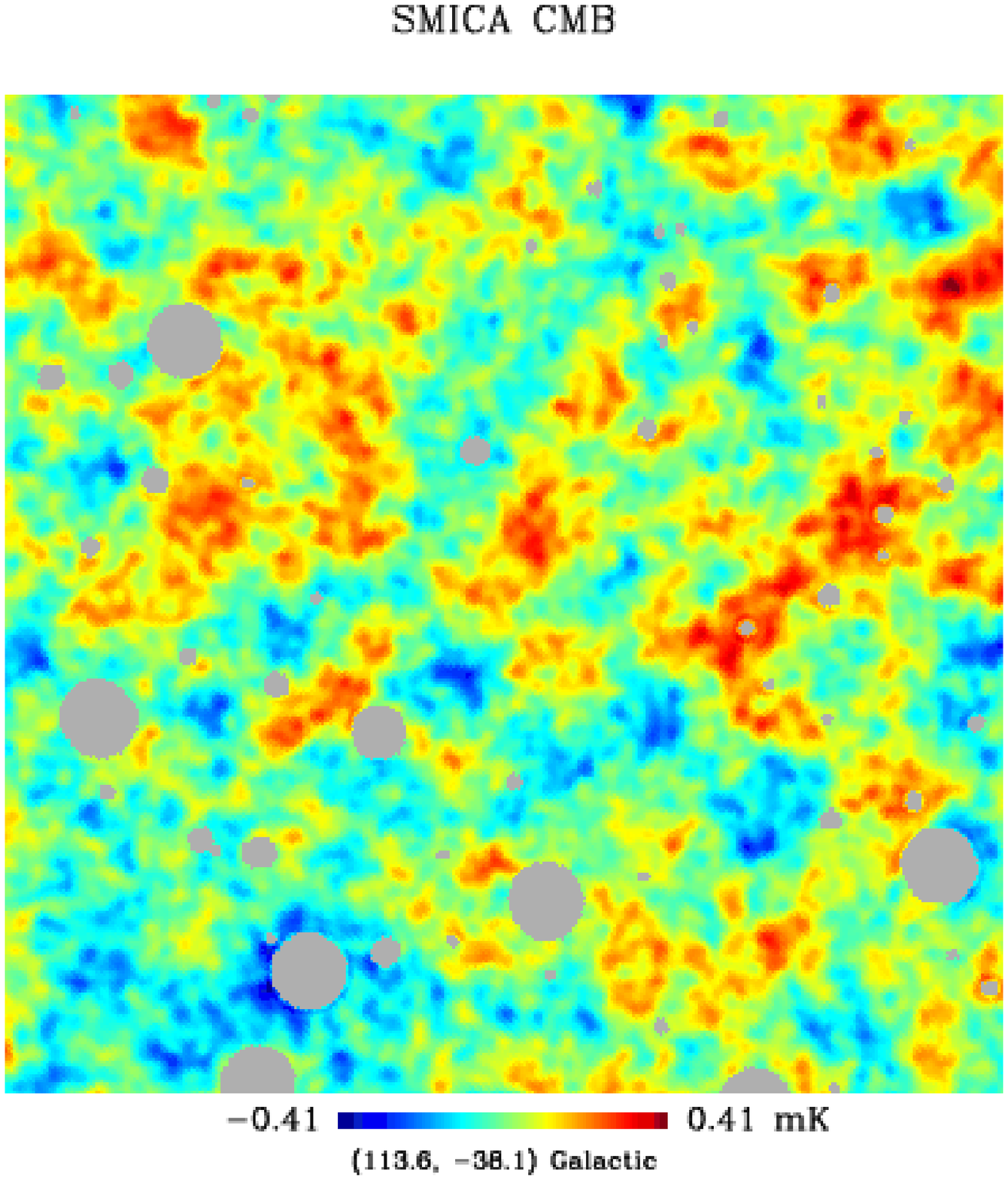}
    \end{tabular}
    \caption[]{{\bf Example input and recovered total galaxy emission at 23GHz, dust at
        143GHz and CMB components.}}
    \label{fig:comp_maps}
    \end{center}
\end{figure*}

In Figure~\ref{fig:comp_maps}, we show for illustration the recovered
total Galactic emission at 23GHz from Commander, the dust emission at
143GHz from FastICA and, for comparison, the recovered CMB from SMICA
on the same patch.

\section{Summary and conclusions}\label{sec:summary}

In this paper we have described a CMB component separation Challenge
based on a set of realistic simulations of the {\sc Planck} satellite
mission. The simulated data were based on a development version of the
{\sc Planck} Sky Model, and included the foreground emission from a
three component Galactic model of free-free, synchrotron and dust, as
well as radio and infra-red sources, the infra-red background, the SZ
effect and {\sc Planck}-like inhomogeneous noise.
We {{\bfref have cautioned}} that the simulations, while complex,
still relied on some simplifying assumptions, {{\bfref as discussed in
Section~\ref{sec:comments}}}.  Thus, there is no guarantee that the
priors and data models that yielded the best separation on simulations
will work equally well on {{\bfref the {\sc Planck} dataset}}.
%
{{\bfref While this may seem to undercut the main purpose of this
    paper, we are simply acknowledging that we cannot anticipate in
    full detail what the {\sc Planck} component separation pipeline
    will look like and how effective it will be, based on an analysis
    of present-day simulations. In spite of this, it is clear that
    methods performing better on the simulated data have in general
    better chances to work better on the real sky.}}

As a combined set of tools, the component separation methods developed
and tested in this work offer very different ways to address the
component separation problem {{\bfref and so} comparable performance
  between different tools, when achieved, provides confidence in the
  conclusions of this work against some of the simplifications used in
  the model for simulating the sky emission.

We found that the recovered CMB maps were clean on large scales, in
the sense that the RMS of the residual contamination was much less
than the cosmic variance: at best the RMS residual of the cleaned CMB
maps was of the order of 2$\mu$K across the sky on a smoothing scale
of $45'$, with a spectral distribution described by $A=0.015
\times\ell^{-0.7} \mu$K$^2$ and $A =0.02\times\ell^{-0.9} \mu$K$^2$ at
high and low Galactic latitudes respectively. The effectiveness of the
foreground removal is illustrated by comparing the input
foreground power spectra of Figure~\ref{fig:clspectra} with the
residuals shown in Figure~\ref{fig:cmbresidualspectra}. The two panels
of the latter figure show that, with few exceptions, the methods
manage to clean the low Galactic latitude Zone~2, where the foreground
contamination is high,
almost as well as they do for the high Galactic latitude Zone~1, where the CMB
dominates at the frequencies near the foreground minimum. The
amplitude of the power spectrum of residuals is, on the largest
scales, four orders of magnitude lower than that of the input Galaxy
power spectrum at the foreground minimum. This means that the CMB map
could be recovered, at least by some methods, over the whole sky
except for a sky cut at the 5 percent level (see
Fig.~\ref{fig:cmbresidualspatial}). The CMB power spectrum was
accurately recovered up to the sixth peak.

As detailed in Table~\ref{tab:methods}, the outputs of the methods
were diverse. While all have produced a CMB map, only a subset of them
were used to obtain maps of individual diffuse Galactic emissions. Five
(Commander, CCA, FastICA, FastMEM, SMICA) reconstructed thermal
dust emission maps at high frequencies, and another five (Commander, CCA,
FastMEM, GMCA, SMICA) yielded a map of the low-frequency Galactic
emissions (synchrotron and free-free).

It is not surprising that the dust component was more easily
reconstructed because it is mapped over a larger frequency range, and
benefits from observations at high frequencies where it dominates over
all other emissions, except the IR background at high Galactic
latitudes. Moreover at high frequencies the noise level is lower and
the angular resolution is better. Low frequency Galactic foregrounds
suffer from more confusion, with a mixture of several components
observed in only few channels, at lower resolution.

The relative errors of the reconstructed foreground maps are larger at
high Galactic latitudes where the foregrounds are fainter. We have
defined a figure of merit $f_{20\%}$, which corresponds to the
fraction of sky where {{\bfref the amplitude of each galactic
component}} has been reconstructed with a relative error of less than
20$\%$. For most methods, $f_{20\%} \simeq 70\%$ was achieved for the dust component,
while $f_{20\%} \simeq 50\%$ was achieved for the radio emission, increasing to
$80\%$ if component separation is performed at a relatively low
resolution of $3^{\circ}$.  Clearly, there is ample room and need for
improvement in this area.

The flux limits for extragalactic point source detection are minimum
at 143 and 217 GHz, where they reach $\simeq 100\,$mJy. About 1000
radio sources and about 2600 far-IR sources are detected over about
67\% of the sky ($|b|>20^\circ$). Over the same region of the sky, the
best methods recover about 2300 clusters.

{{\bfref In closing, we list areas where work is in progress and
    improvements are expected:}}
The sky model is being upgraded to include the anomalous emission
component and polarization. We are in the process of integrating point
source and SZ extraction algorithms together with the diffuse
component separation algorithms into a single component separation
pipeline.  This is expected, on one side, to decrease the
contamination of CMB maps on small angular scales, where point and
compact sources (including SZ effects) dominate and, on the other
side, to achieve a more efficient point and compact source extraction.
{{\bfref Assessing this in more detail is part of WG2 work plans for
    the near future. The SEVEM foreground cleaning step now operates in
    wavelet space which allows for improved, scale-by-scale removal of
    foregrounds. In addition the recovery of the power spectrum
    estimates and error bars at the highest multipoles has been
    improved by reducing the cross-correlation between modes through
    the use of an apodised mask. For Commander, work is currently
    ongoing to extend the foreground sampler to multi-resolution
    experiments. CCA is being upgraded to fully exploit the estimated
    spatially-varying spectral indices in the source reconstruction step;
    SMICA is being improved to model unresolved point
    source power. The FastICA algorithm is being improved to handle
    data with a wider frequency coverage. The GMCA framework is being
    extended to perform a joint separation and deconvolution of the
    components.  }}

\begin{acknowledgements}
  The work reported in this paper was carried out by Working Group 2 of
  the {\sc Planck} Collaboration.
  {\sc Planck} is a mission of the  European Space Agency.                                                              
  The Italian group were supported in part by
  ASI (contract Planck LFI Phase E2 Activity). 
  The US {\sc Planck} Project is supported by the NASA
  Science Mission Directorate.
  The ADAMIS team at APC has been partly supported by the Astro-Map and Cosmostat ACI grants
  of the French ministry of research, for the development of innovative CMB data analysis methods.
  The IFCA team acknowledges partial financial support from the Spanish project
  AYA2007-68058-C03-02
  This research used resources of the National Energy Research
  Scientific Computing Center, which is supported by the Office of
  Science of the U.S. Department of Energy under Contract No.
  DE-AC02-05CH11231.
  DH, JLS and ES would like to acknowledge partial financial support from joint CNR-CSIC research project
  2006-IT-0037.
  RS acknowledges support of the EC Marie Curie IR Grant (mirg-ct-2006-036614).
  CB was partly supported by the NASA LTSA Grant NNG04CG90G.
  SML would like to thank the kind hospitality of the APC, Paris for
  a research visit where part of this work was completed.
  We thank Carlo Burigana for a careful reading of the manuscript.
  We acknowledge the use of the Legacy Archive for Microwave Background Data
  Analysis (LAMBDA). Support for LAMBDA is provided by the NASA Office of Space Science.
  The results in this paper have been derived using the
  HEALPix package \citep{2005ApJ...622..759G}.
\end{acknowledgements}

\bibliographystyle{aa} 
\bibliography{compsepcomp} 
\appendix

\section{Description of methods}\label{sec:methods}

\subsection{Commander}

`Commander' is an implementation of the CMB and foregrounds
Gibbs sampler most recently described by \citet{2008ApJ...676...10E}; This
algorithm maps out the joint CMB-foreground probability distribution, or
`posterior distribution', by sampling.
The target posterior distribution may be written in terms of the likelihood and
prior using Bayes' theorem,
\begin{eqnarray}
  \Pr(\Bs, C_{\ell}, \theta_{\textrm{fg}}|\Bd) &= \mathcal{L}(\Bd|\Bs,
  \theta_{\textrm{fg}}) \Pr(\Bs|C_{\ell}) \Pr(C_{\ell})
  \Pr(\theta_{\textrm{fg}}).
\label{eq:posterior}
\end{eqnarray}
Here 
$\theta_{\textrm{fg}}$ is the collection of all parameters required to
describe the non-cosmological foregrounds. Since the noise is assumed
to be Gaussian, the likelihood is simply given by the $\chi^2$.

In the current analysis, the foregrounds are modelled by a sum of
synchrotron, free-free and thermal dust emission, and free monopole
and dipoles at each frequency band. The thermal dust component is
approximated by a single-component modified blackbody with a fixed
dust temperature $T_d = 21\textrm{K}$. Thus, the total foreground
model reads
\begin{eqnarray}
s_{\rm fg}(\nu,p) &=&  A_s(p) g(\nu) \left(\frac{\nu}{\nu_s}\right)^{\beta_s(p)}
+ A_{f}(p) g(\nu) \left(\frac{\nu}{\nu_{f}}\right)^{-2.15} \nonumber \\
&&+A_{d}(p) g(\nu) \frac{e^{h\nu_d/kT_d}-1}{e^{h\nu/kT_d}-1}\left(\frac{\nu}{\nu_{d}}\right)^{\beta_d(p)+1}\nonumber
\\
&&+m^0_{\nu} + \sum_{i=1}^{3} m^i_{\nu} (\hat{\mathbf{n}}(p) \cdot \hat{\mathbf{e}}_i),
\end{eqnarray}
where $g(\nu)$ is the conversion factor between antenna and
thermodynamic temperatures, and $\hat{\mathbf{n}}$ is the unit vector
of pixel $p$. The free parameters are thus the foreground amplitudes,
$A_s$, $A_f$ and $A_d$, and spectral indices, $\beta_s$ and $\beta_d$,
for each pixel, and the overall monopole, $m^0_{\nu}$, and dipole
amplitudes, $m^i_{\nu}$, for each band. For priors, we adopt the
product of the Jeffreys' ignorance prior and an informative Gaussian
prior ($\beta_s = -3 \pm 0.3$ for synchrotron and $\beta_d = 1.5 \pm
0.3$ for dust) for the spectral indices, while no constraints are
imposed on the amplitudes.

Using the Gibbs (conditional) sampling technique, a set of samples drawn from the
posterior distribution. From these samples the marginal posterior mean and
RMS component maps are derived, as well as the marginal CMB power spectrum
posterior distribution.

The code assumes identical beams at all
frequencies, and it is therefore necessary to smooth the data to a
common resolution, limiting the analysis to large angular
scales. For this particular data set, we have chosen a common
resolution of $3^{\circ}$ FWHM, with $54'$ pixels (Healpix $N_{\textrm{side}}$=64)
and with $\ell_{\textrm{max}}=150$. For more details on the
degradation process, see \citet{2008ApJ...676...10E}. At this resolution,
the CPU time for producing one sample is around one wall-clock minute. A
total of 5400 samples were produced over four independent Markov
chains, of which the first 2400 were rejected due to burn-in. Twelve
frequency bands (covering frequencies between 23 and 353 GHz) were
included, for a total cost of around 1000 CPU hours.

The main advantage of this approach is simply that it provides us with
the exact joint CMB and foreground posterior distribution for very general
foreground models. From this joint posterior distribution, it is trivial to obtain
the exact marginal CMB power spectrum and sky signal
posterior distributions. Second, since any parametric foreground model may be
included in the analysis, the method is very general and flexible. It
also provides maps of the posterior means for individual components, and is
therefore a true component separation method, and not only a
foreground removal tool.

Currently, the main disadvantage of the approach is the assumption of
identical beam profiles at each frequency. This strictly limits the
analysis to the lowest resolution of a particular data set. However,
this is a limitation of the current implementation, and not of the
method as such.

\subsection{Correlated Component Analysis (CCA)}

CCA \citep{ccapisa} is a semi-blind approach that relies on the
second-order statistics of the data to estimate the mixing matrix
on sub-patches of the sky.
CCA assumes the data model given by Eq.~(\ref{eq:mixing}), 
and makes no assumptions about the independence or lack of
correlations between pairs of radiation sources.
The method exploits the spatial structure of the individual source
maps and adopts commonly accepted models for source frequency
scalings in order to reduce the number of free parameters to be estimated.

The spatial structures of the maps are accounted for through the
covariance matrices at different shifts.  From the data model adopted,
the data covariance matrices at shifts $(\tau,\psi)$ are given by
\begin{eqnarray} \label{eq:cov_shift0}
  \mathbf{C}_{\rm d}(\mathrm{\tau},\psi) & = & \langle \left[ \mathbf{d}(\theta,\phi) -
    \mu_{\rm d} \right]
  \left[ \mathbf{d}(\theta+\tau,\phi+\psi)-\mu_{\rm d} \right]^{\mathrm{t}} \rangle
  \nonumber \\
  &= &\mathbf{AC}_{\rm s}(\mathrm{\tau},\psi)\mathbf{A}^{\mathrm{t}}+\mathbf{C}_{\rm n}(\tau,\psi)
  \ .  
\end{eqnarray}
where $\mu_{\rm d}$ is the mean data vector, and $(\theta,\phi)$ is the
generic pixel index pair. The matrices
$\mathbf{C}_{\rm d}(\mathrm{\tau},\psi)$ can be estimated from the data,
and the noise covariance matrices $\mathbf{C}_{\rm n}(\tau,\psi)$ are
derived from the map-making noise estimations. From
Eq.~(\ref{eq:cov_shift0}), we can estimate the mixing matrix and free
parameters of the source covariance matrices by matching the known
quantities to the unknowns, that is by minimizing the following
function for $\mathbf{A}$ and $\mathbf{C}_{\rm s}(\mathrm{\tau},\psi)$
\begin{eqnarray} \label{eq:estimation_acs}
\sum_{\tau,\psi}^{}\|\mathbf{A}\mathbf{C}_{\rm s}(\tau,\psi)\mathbf{A}^{\mathrm{t}}-
[\mathbf{C}_{\rm d}(\tau,\psi)-\mathbf{C}_{\rm n}(\tau,\psi)]\| ,
\end{eqnarray}
where the Frobenius norm is used and the summation is taken over the
set of shift pairs $(\tau,\psi)$ for which data covariances are
non-zero. Given an estimate of $\mathbf{C}_{\rm s}$ and $\mathbf{C}_{\rm n}$,
Eq.~(\ref{eq:mixing}) can be inverted and component maps obtained via
the standard inversion techniques of Wiener filtering or generalised
least square inversion. For the Challenge, harmonic space Wiener
filtering was applied, using a mixing matrix obtained by averaging the
mixing matrices of different patches. More details on the method can
be found in \citet{2006MNRAS.373..271B,2007MNRAS.382.1791B}.

CCA can treat the variability of the spectral properties of each
component with the direction of observation by working on sufficiently
small sky patches, which must however be large enough to have
sufficient constraining power; typically the number of pixels per
patch must be around $10^5$. To obtain a continuous distribution of
the free parameters of the mixing matrix, CCA is applied to a large
number of partially overlapping patches.

A drawback of the present version of CCA is common to many
pixel-domain approaches to separation: the data must be smoothed
to a common resolution. A Fourier-domain implementation of
CCA \citep{bedini2007} would be able to cope with this problem.
Alternatively for the pixel-domain version, the mixing matrix could be 
estimated from the smoothed maps and then used to separate the sources
using the full resolution data.

\subsection{Generalised morphological component analysis (GMCA)}



GMCA \citep{gmca} is a blind source separation method devised for
separating sources from instantaneous linear mixtures using the model
given by Eq.~(\ref{eq:mixing}).
The components $\bf s$ are assumed to be sparsely represented
(i.e. have a few significant samples in a specific basis) in
a so-called sparse representation $\bf \Phi$ (typically
wavelets). Assuming that the components have a sparse representation
in the wavelet domain is equivalent to assuming that most components
have a certain spatial regularity. These components and their spectral
signatures are then recovered by minimizing the number of significant
coefficients in $\bf \Phi$~:
\begin{eqnarray}
\mbox{min}_{\{{{\bf a},{\bf s}}\}} \lambda \|{\bf s}{\bf \Phi}^T\| + \frac{1}{2}\|{\bf d} - {\bf as}\|_2^2
\end{eqnarray}
In \cite{gmca}, it was shown that sparsity enhances the
diversity between the components thus improving the separation
quality. The spectral signatures of CMB and SZ are assumed to be
known. The spectral signature of the free-free component is
approximately known up to a multiplicative constant (power law with
fixed spectral index). The synchrotron component is estimated via a
separable linear model~: ${\bf d}_{sync} = a_{sync} s_{sync}$ where
$a_{sync}$ is parameterised by a spectral index $\beta_{sync}$. This
spectral index is estimated by solving the following
problem:
\begin{eqnarray}
\min_{\beta} \|{\bf r}_{\rm sync} - a_{sync}(\beta) s_{\rm Haslam}\|_2^2
\end{eqnarray}
where ${\bf r}_{\rm sync}$ is the residual obtained by extracting the
contribution of all the components from the data $\bf d$ except
synchrotron. $s_{\rm Haslam}$ is the Haslam synchrotron map;
$a_{sync}(\beta)$ is the spectral signature of synchrotron emission
(power law). More precisely, $\beta$ is estimated such that the Haslam
multiplied by $a_{sync}(\beta)$ matches the residual term ${\bf
r}_{sync}$.

A Wiener filter is applied to provide the denoised CMB estimate.  The
main advantage of GMCA is its ability to blindly extract strong
galactic emission. Indeed, most galactic emission is well represented
in a wavelet basis. The main disadvantage is that it relies on the way
the deconvolution of the data is performed: an effective beam is used
to account for the convolution.

\subsection{Independent component analysis (FastICA)}

Independent Component Analysis is an approach to component
separation, looking for the components which maximise some 
measure of the statistical independence \citep{fastica}. 
The FastICA algorithm presented here exploits the fact that 
non-Gaussianity is usually a convenient and robust measure of 
the statistical independence and therefore it searches for 
linear combinations ${\bf y}$ of the input multi-frequency data, 
which maximise some measure of the non-Gaussianity. 
In the specific implementation of the idea, employed here, 
the non-Gaussianity is quantified by the {\em neg-entropy}. 
Denoting by $H({\bf y})=-\int p({\bf y})\log p({\bf y})d{\bf y}$ 
the entropy associated with the distribution $p$, we define 
the neg-entropy as,
\begin{eqnarray}
\label{neg-entropy}
\textrm{neg-entropy}({\bf y})=H({\bf y}_{G})-H({\bf y})\ ,
\end{eqnarray}
where ${\bf y}_{G}$ is a Gaussian variable with the same covariance
matrix as ${\bf y}$. The search for the maxima of the neg-entropy is
usually aided by enhancing the role of the higher order moments of
${\bf y}$, which is achieved by means of a non-linear mapping. In the
present implementation, the FastICA finds the extrema of the
neg-entropy approximation given by $|E[g({\bf y})]-E[g({\bf
y_{G}})]|^{2}$, where $E$ means the average over the pixels, and $g$
represents the non-linear mapping of the data, which may be a power
law in the simplest case.  The algorithm is straightforwardly
implemented in real space, and requires the same angular resolution
for all channels.  Note that for an experiment like {\sc Planck} where the
resolution varies with frequency, this requires smoothing the
input data to the lowest resolution before processing. The use of an
efficient minimization procedure, with a required number of floating
point operations scaling linearly with the size of the data set, makes
the computational requirements essentially dominated by memory needed
to be allocated to quickly access the multi-frequency data.

The algorithm has been tested so far as a CMB cleaning procedure, because
the hypothesis of statistical independence is expected to be verified
at least between CMB and diffuse foregrounds.  It produced results on
real (BEAST, COBE, WMAP) and simulated total intensity data, as well
as on polarization simulations, on patches as well as all sky (see
\citet{2007MNRAS.374.1207M} and references therein). The performance is made
possible by two contingencies, i.e. the validity of the assumption of
statistical independence for CMB and foregrounds, as well as the high
resolution of the present CMB observations, which provides enough of
statistical realizations (pixels) for the method to decompose the data
into the independent components.

\subsection{Harmonic-space maximum entropy method (FastMEM)}

The Maximum Entropy Method (MEM) can be used to separate the CMB
signal from astrophysical foregrounds including Galactic synchrotron,
dust and free-free emission as well as SZ effects.  The particular
implementation of MEM used here works in the spherical harmonic
domain.  The separation is performed mode-by-mode allowing  a
huge optimisation problem to be split into a number of smaller problems.  The
solution can thus be obtained more rapidly, giving this implementation its
name: FastMEM.  This approach is described by
\citet{1998MNRAS.300....1H,1999MNRAS.306..232H}
for Fourier modes on flat patches of the sky and by \citet{mem,Stolyarov:2004xp}
for the full-sky case.

If we have a model (or hypothesis) $H$ in which the measured data $\mathbf{d}$
is a function of an underlying signal $\mathbf{s}$, then Bayes' theorem tells
us that the posterior probability $\Pr(s|d,H)$ is the product of the
likelihood $\Pr(\mathbf{d}|\mathbf{s},H)$ and the prior probability $\Pr(\mathbf{s},H)$, divided
by the evidence $\Pr(\mathbf{d},H)$,
\begin{eqnarray}
\Pr(\mathbf{s}|\mathbf{d},H) = \frac{\Pr(\mathbf{d}|\mathbf{s},H)\Pr(\mathbf{s}|H)}{\Pr(\mathbf{d}|H)}.
\end{eqnarray}
The objective here is to maximise the posterior probability of the
signal given the data. Since the evidence in Bayes' theorem is merely
a normalisation constant we maximise the product of the likelihood and
the prior
\begin{eqnarray}
\Pr(\mathbf{s}|\mathbf{d},H) \propto \Pr(\mathbf{d}|\mathbf{s},H) \Pr(\mathbf{s},H).
\end{eqnarray}
We assume that the instrumental noise in each frequency channel is
Gaussian-distributed, so that the log-likelihood has a form of a
$\chi^2$ misfit statistic.  We make the assumption that the noise is
uncorrelated between spherical harmonic modes.  We also assume that
the beams are azimuthally symmetric, so that they are fully described
by the beam transfer function $B_\ell$ in harmonic space.  For mode
$(\ell,m)$, the log-likelihood is
\begin{eqnarray}
  \chi^2(\mathbf{s}_{\ell m})  = 
  (\mathbf{d}_{\ell m} - B_\ell \mathbf{A} \mathbf{s}_{\ell m})^T
  N^{-1}_{\ell m}
  (\mathbf{d}_{\ell m}- B_\ell \mathbf{A} \mathbf{s}_{\ell m})
\end{eqnarray}
where $\mathbf{A}$ is the fixed frequency conversion matrix which describes how
the components are mixed to form the data, and $N^{-1}_{\ell m}$ is the
inverse noise covariance matrix for this mode.  If the instrumental
noise is uncorrelated between channels, then this matrix is diagonal.
However, unresolved point sources can be modelled as a correlated
noise component.

The prior can be Gaussian, and in this case we recover the Wiener
filter with the well-known analytical solution for the signal $s$.
However, the astrophysical components have strongly non-Gaussian
distribution, especially in the Galactic plane.  Therefore
\citet{1998MNRAS.300....1H} suggested that an entropic prior be used
instead. In this case, maximising the posterior probability is equivalent to the
minimising the following functional for each spherical harmonic mode
\begin{eqnarray}
\Phi_{\rm MEM}(\mathbf{s}_{\ell m}) = \chi^2(\mathbf{s}_{\ell m})-\alpha S(\mathbf{s}_{\ell m})
\end{eqnarray}
%
where $S(\mathbf{s})$ is the entropic term, and $\alpha$ is the regularisation
parameter.  The minimisation can be done numerically using one of a
number of algorithms \citep{1992nrfa.book.....P}.

FastMEM is a non-blind method, so the spectral behaviour of the
components must be known in advance.  Since $\mathbf{A}$ is fixed, the spectral
properties of the components must be the same everywhere on the sky.
However, small variations in the spectral properties, for example,
dust temperature, synchrotron spectral index or SZ cluster electron
temperature, can be accounted for by introducing additional
components.  These additional components correspond to terms in the
Taylor expansion of the frequency spectrum with respect to the
relevant parameter.

The initial priors on the components are quite flexible and they can
be updated by iterating the component separation, especially if the
signal-to-noise is high enough.

It is not necessary for all of the input maps to be at the same
resolution since FastMEM solves for the most probable solution for
unsmoothed signal, deconvolving and denoising maps simultaneously. It
is flexible enough to include any datasets with known window function
and noise properties.  A mask can easily be applied to the input data
(the same mask for all frequency channels) and this does not cause
problems with the separation.

Since FastMEM uses priors on the signals, the solution for the signals
is biased.  This is especially evident if the signal-to-noise ratio is
low.  It is possible to de-bias the power spectrum statistically,
knowing the priors and the FastMEM separation errors per mode.
However, one can not de-bias the recovered maps since the errors are
quadratic and de-biasing will introduce phase errors in the harmonics.

No information about the input components was used in the separation,  
and the prior power spectra were based solely on the physical  
properties of the components and templates available in the  
literature. The prior on the CMB component was set using the best-fit  
theoretical spectrum, instead of a WMAP--constrained realisation. This  
has a significant effect at low multipoles.

\subsection{Spectral estimation via expectation-maximization (SEVEM)}

SEVEM \citep{Martinez-Gonzalez:2003} tries to recover
only the CMB signal, treating the rest of the emissions as a
generalised noise. As a first step, the cosmological frequency maps,
100, 143 and 217 GHz, are foreground cleaned using an internal
template fitting technique. Four templates are obtained from the
difference of two consecutive frequency channels, which are smoothed
down to the same resolution if necessary,to avoid the presence of
CMB signal in the templates. In particular, we construct maps of
(30-44), (44-70), (545-353) and (857-545) differences. The central frequency
channels are then cleaned by subtracting a linear combination of these
templates. The coefficients of this combination are obtained
minimising the variance of the final clean map outside the considered
mask. The second step consists of estimating the power spectrum of
the CMB from the three cleaned maps using the method (based on the
Expectation-Maximization algorithm) described in
\citet{Martinez-Gonzalez:2003}, which has been adapted to deal
with spherical data. Using simulations of CMB plus noise, processed in
the same way as the Challenge data, we obtain the bias and statistical
error of the estimated power spectrum and construct an unbiased
version of the $C_{\ell}$'s of the CMB. This unbiased power spectrum
is used to recover the CMB map from the three clean channels through
Wiener filter in harmonic space. Finally, we estimate the noise per
pixel of the reconstructed map using CMB plus noise simulations.

One of the advantages of SEVEM is that it does not need any external
data set or need to make any assumptions about the frequency dependence or
the power spectra of the foregrounds, other than the fact that they
are the dominant contribution at the lowest and highest frequency
channels. This makes the method very robust and, therefore, it is
expected to perform well for real {\sc Planck} data. Moreover, SEVEM
provides a good recovery of the power spectrum up to relatively high
$\ell$ and a small error in the CMB map reconstruction. In addition
the method is very fast, which allows one to characterise the errors
of the CMB power spectrum and map using simulations. The cleaning of
the data takes around 20 minutes, while the estimation of the power
spectrum and map requires around 15 and 30 minutes respectively. In
fact, the whole process described, including producing simulations to
estimate the bias and errors, takes around 30 hours on one single CPU.
Regarding weak points, the method reconstructs only the CMB
and does not try to recover any other component of the microwave sky
although it could be generalised to reconstruct simultaneously the
both the CMB and the thermal SZ effect. Also, the reconstructed CMB map is not
full-sky, since the method does not aim to remove the strong
contamination at the centre of the Galactic plane or at the point
source positions. In any case, the masked region excluded for the
analysis is relatively small.

\subsection{Spectral matching independent component analysis (SMICA)}

The principle of SMICA can be summarised in three steps: 
1) Compute spectral statistics. 
2) Fit a component-based model to them. 
3) Use the result to implement a Wiener filter in harmonic space.
More specifically, an idealised operation goes as follows.  
Denote $\mathbf{d}(\xi)$ the column vector whose $i$-th entry contains
the observation in direction $\xi$ for the $i$-th channel and denote
$\mathbf{d}_{\ell m}$ the vector of same size (the number of frequency
channels) in harmonic space.
This is modelled as the superposition of $C$ components
$\mathbf{d}_{\ell m} = \sum_{c=1}^C\mathbf{d}_{\ell m}^c$.
In Step 1), we compute spectral matrices $\widehat \mathbf{C}_\ell =
\frac1{2\ell+1}\sum_m \mathbf{d}_{\ell m}^{ }  \mathbf{d}_{\ell m}^{\mathrm{T}}$.
In Step 2) we model the ensemble-averaged spectral matrix 
$\mathbf{C}_\ell = \langle \widehat \mathbf{C}_\ell\rangle$ as the superposition of $C$
uncorrelated components: 
$\mathbf{C}_\ell = \sum_{c=1}^C \mathbf{C}_\ell^c$ and, for each component, we postulate
a parametric model, that is, we let the matrix set
$\{\mathbf{C}_\ell^c\}_{\ell=0}^{\ell_\mathrm{max}}$ be a function of a
parameter vector $\theta^c$. 
This parameterization embodies our prior knowledge about a given
component.  For instance, for the CMB component, we take
$[\mathbf{C}_\ell^\mathrm{cmb}]_{ij} = e_i e_j c_\ell$ where $e_i$ is the known CMB
emmission coefficient for channel $i$ and $c_\ell$ is the unknown
angular power spectrum at frequency~$\ell$.  The parameter vector for CMB
would then be
$\theta^\mathrm{cmb}=\{c_\ell\}_{\ell=0}^{\ell_\mathrm{max}}$.
All unknown parameters for all components are then estimated  by fitting the
model to the spectral statistics, i.e. by solving
$ 
  \min_{\theta^1,\ldots,\theta^C}
  \sum_{\ell=0}^{\ell_\mathrm{max}} (2\ell+1)
  \ K[ \ \widehat \mathbf{C}_\ell\ |\ \sum_{c=0}^C \mathbf{C}_\ell^c(\theta^c) \ ]
$
where $K[\mathbf{C}_1|\mathbf{C}_2]$ is a measure of mismatch between two covariance
matrices $\mathbf{C}_1$ and $\mathbf{C}_2$.  The resulting values
$\hat\theta^1,\ldots,\hat\theta^C$ provide estimates
$\mathbf{C}_\ell^c(\hat\theta^c)$ of $\mathbf{C}_\ell^c$.
The Wiener filter estimate of $\mathbf{d}^c_{\ell m}$ can be expressed
as $\hat{\mathbf{d}}^{c}_{\ell m} = \mathbf{C}_\ell^c \mathbf{C}_\ell^{-1} \mathbf{d}_{\ell
  m}$.  In practice, we use the fitted spectral matrices estimated at
the previous step: component $c$ is estimated as 
\begin{eqnarray}
  \hat{\mathbf{d}}^c_{\ell m}   =
  \mathbf{C}_\ell^c(\hat\theta^c) \mathbf{C}_\ell(\hat\theta)^{-1} \mathbf{d}_{\ell m}
\end{eqnarray}
and the maps of each component in each channel are finally computed by
inverse spherical harmonic transforms.

For processing the current data set, we have used a model containing
four components: the CMB, the SZ component, a 4-dimensional Galactic
component and a noise component.

The actual processing includes several modifications with respect to
this outline: a) beam correction applied to each spectral matrix
$\widehat \mathbf{C}_\ell$; b) spectral binning by which the (beam corrected)
spectral matrices are averaged over bins of increasing lengths; c)
localization implemented via aopdised masks, by which the SMICA
process is conducted independently over two different sky zones.

Strengths: a) No prior information used regarding Galactic emission. b)
Accurate recovery of the CMB via Wiener filtering. c) It is a relatively fast
algorithm. d) Built-in goodness of fit.

Weaknesses: a) The results reported here do not account for the
contribution of point sources for which a convenient model is lacking.
b) Localization in two zones is probably too crude. c) No separation of
Galactic components.

\subsection{Wavelet-based hIgh-resolution Fitting of Internal Templates (WI-FIT)}

WI-FIT \citep{Hansen:2006rj} is based on fitting and subtraction of
internal templates. Regular (external) template fitting uses external
templates of Galactic components based on observations at frequencies
different from the ones used to study the CMB. These templates are
fitted to CMB data, the best fit coefficients for each component are
found and the templates are subtracted from the map using these
coefficients in order to obtain a clean CMB map. WI-FIT differs from
this procedure in two respects: (1) It does not rely on external
observations of the galaxy but forms templates by taking the
difference of CMB maps at different channels. The CMB temperature is
equal at different frequencies whereas the Galactic components are
not. For this reason, the difference maps contain only a sum of
Galactic components. A set of templates are constructed from
difference maps based on different combinations of channels. (2) The
fitting of the templates are done in wavelet space where the
uncertainty on the foreground coefficients is much lower than a
similar pixel based approach (in the pixel based approach, no
pixel-pixel correlations are taken into account since the correlation
matrix will become to large for {\sc Planck}-like data sets. In the wavelet
based approach, a large part of these correlations are taken into
account in scale-scale covariance matrices).

For calibration purposes, a set of 500 simulated CMB maps need to be
produced and the full wavelet fitting procedure applied to all maps.
This is where most CPU time goes. For {\sc Planck} resolution maps,
around $1$ Gb of memory is necessary to apply WI-FIT and a total of around $400$
CPU hours are required.

The strength of WI-FIT is that it relies on very few assumptions about
the Galactic components. WI-FIT does however assume that the spectral
indices do not vary strongly from pixel to pixel within the frequency
range used in the analysis. If this assumption is wrong then WI-FIT leaves
residuals in the areas where there are strongly varying spectral indices.

Another advantage of WI-FIT is that it is easy to apply and is
completely linear, i.e. the resulting map is a linear combination of
frequency channels with well known noise and beam properties. This
will in general result in increased noise variance in the cleaned
map. In order to avoid this, we smooth the internal templates in order
to make the noise at small scales negligable and at the same time not
make significant changes to the shape of the diffuse foregrounds. If
the diffuse foregrounds turn out to be important at small scales
$l>300$, the smoothing of the internal templates will significantly
reduce the ability of WI-FIT to perform foreground cleaning at these
scales. Tests on the {\sc WMAP} data have shown that diffuse foregrounds do
not seem to play an important role at such small scales. This is valid
for the frequency range observed by {\sc WMAP} (i.e. at LFI-frequencies),
similar tests will need to be made for the {\sc Planck} HFI data.

Finally, WI-FIT does not do anything to the point sources, which  need
to be masked.

\end{document}